\documentclass[proceedings]{JHEP3}
\usepackage{amsfonts}
\usepackage{amsmath}
\usepackage{epsfig}

\newbox\mybox

\newcommand\fverb{\setbox\mybox=\hbox\bgroup\verb}
\newcommand\fverbdo{\egroup\medskip\noindent\fbox{\unhbox\mybox}\ }
\newcommand\fverbit{\egroup\item[\fbox{\unhbox\mybox}]}
\conference{Quantum mechanics in time-dependent backgrounds}
\abstract{We investigate a quantum mechanical system 
on a noncommutative space for which the structure constant is explicitly 
time-dependent. Any autonomous Hamiltonian on such a space acquires 
a time-dependent form in terms of the conventional canonical variables. We 
employ the Lewis-Riesenfeld method of invariants to construct explicit
analytical solutions for the corresponding time-dependent Schr\"{o}dinger equation. 
The eigenfunctions are expressed in terms of the solutions of variants of the nonlinear Ermakov-Pinney 
equation and discussed in detail for various types of background fields.
We utilize the solutions to verify a generalized version of
Heisenberg's uncertainty relations for which the lower bound becomes a time-dependent
function of the background fields. We study the variance for various states including
standard Glauber coherent states with their squeezed versions and Gaussian Klauder coherent states
resembling a quasi-classical behaviour. No type of coherent states appears to be optimal in general with regard to achieving minimal uncertainties, as this feature turns out to be 
background field dependent.}

\title{Noncommutative quantum mechanics in a time-dependent background}
\author{Sanjib Dey and Andreas Fring \\
Department of Mathematics, City University London,\\
Northampton Square, London EC1V 0HB, UK\\
E-mail: sanjib.dey.1@city.ac.uk, a.fring@city.ac.uk}

\input{tcilatex}
\begin{document}

\section{Introduction}

The study of quantum mechanics and quantum field theories on noncommutative
space-time structures is motivated by the fact that it achieves
gravitational stability \cite{dop} in almost all currently known approaches
to quantum gravity, such as string theory \cite{String1,String2,wit} or loop
quantum gravity \cite{Ashtekar,Rovelli}. In a quantum mechanical setting the
most commonly studied version of these space-time structures consists of
replacing the standard set of commutation relations for the canonical
coordinates $x^{\mu }$ by noncommutative versions, such as $[x^{\mu
},x^{v}]=i\theta ^{\mu \nu }$, where $\theta ^{\mu \nu }$ is taken to be a
constant antisymmetric tensor. More interesting structures, leading for
instance to minimal length and generalized versions of Heisenberg's
uncertainty relations, are obtained when $\theta ^{\mu \nu }$ is taken to be
a function of the momenta and coordinates, e.g. \cite%
{Kempf2,AFBB,AFLGBB,DFG,DF2}. In addition, one may of course also introduce
an explicit time-dependence in $\theta ^{\mu \nu }$. Although various
effective Lagrangians for such type of noncommutative field theories have
been derived, e.g. \cite{Calmet}, little is known about explicit quantum
theories on such type of spaces, one of the reasons being that they are far
more difficult to solve.

Here our aim is to find explicit solutions for a simple prototype quantum
mechanical model on a time-dependent background and study the physical
consequences such a space will imply. We focus here on the particular
two-dimensional space with nonvanishing commutators for the coordinates $X$, 
$Y$ and momenta $P_{x},P_{y}$ 
\begin{equation}
\left[ X,Y\right] =i\theta (t),\qquad \left[ P_{x},P_{y}\right] =i\Omega
(t),\qquad \left[ X,P_{x}\right] =\left[ Y,P_{y}\right] =i\hbar +i\frac{%
\theta (t)\Omega (t)}{4\hbar },  \label{space}
\end{equation}%
where the noncommutative structure constants $\theta (t)$ and $\Omega (t)$
are taken to be real valued functions of time $t$. Of course a multitude of
other possibilities exists. As we shall see in detail below, when
considering representations for these phase-space variables one is
inevitably lead to time-dependent Hamiltonians $H(X,Y,P_{x},P_{y})%
\rightarrow H(t)$.

We will employ here the method of invariants, introduced originally by Lewis
and Riesenfeld \cite{Lewis69}, to solve the time-dependent Schr\"{o}dinger
equation 
\begin{equation}
i\hbar \partial _{t}\left\vert \psi _{n}\right\rangle =H(t)\left\vert \psi
_{n}\right\rangle ,  \label{SCH}
\end{equation}%
for the time-dependent or dressed states $\left\vert \psi _{n}\right\rangle $
associated to the Hamiltonian $H(t)$.

Let us briefly describe the key steps of the method for future reference.
The initial step in that approach consists of constructing a Hermitian
time-dependent invariant $I(t)$ from the evolution equation%
\begin{equation}
\frac{dI(t)}{dt}=\partial _{t}I(t)+\frac{1}{i\hbar }[I(t),H(t)]=0.
\label{Inv}
\end{equation}%
In the next step one needs to solve the corresponding eigenvalue system
involving the invariant%
\begin{equation}
I(t)\left\vert \phi _{n}\right\rangle =\lambda \left\vert \phi
_{n}\right\rangle ,  \label{eigen}
\end{equation}%
for real and time-independent eigenvalues $\lambda $ and for time-dependent
states $\left\vert \phi _{n}\right\rangle $. It was shown in \cite{Lewis69}
that the states%
\begin{equation}
\left\vert \psi _{n}\right\rangle =e^{i\alpha (t)}\left\vert \phi
_{n}\right\rangle  \label{ph}
\end{equation}%
satisfy the time-dependent Schr\"{o}dinger equation (\ref{SCH}) provided
that the real function $\alpha (t)$ in (\ref{ph}) obeys%
\begin{equation}
\frac{d\alpha (t)}{dt}=\frac{1}{\hbar }\left\langle \phi _{n}\right\vert
i\hbar \partial _{t}-H(t)\left\vert \phi _{n}\right\rangle .  \label{pht}
\end{equation}%
For more details on the derivation of these key equations we refer the
reader to \cite{Lewis69}.

Having obtained the explicit solutions for the wavefunctions one is in the
position to compute expectation values for any desired observable. Of
special interest is to investigate the modified version of Heisenberg's
uncertainty relations resulting from non-vanishing commutation relations (%
\ref{space}). Following standard arguments, the uncertainty for the
simultaneous measurement of the observables $A$ and $B$ has to obey the
inequality%
\begin{equation}
\left. \Delta A\Delta B\right\vert _{\psi }\geq \frac{1}{2}\left\vert
\left\langle \psi \right\vert [A,B]\left\vert \psi \right\rangle \right\vert
,  \label{GHU}
\end{equation}%
with $\left. \Delta A\right\vert _{\psi }^{2}=\left\langle \psi \right\vert
A^{2}\left\vert \psi \right\rangle -\left\langle \psi \right\vert
A\left\vert \psi \right\rangle ^{2}$ and similarly for $B$ for any state $%
\left\vert \psi \right\rangle $. Evidently, for instance the first relation
in (\ref{space}) implies that the uncertainty for the simultaneous
measurement of $X$ and $Y$ is greater than the function of time $\left\vert
\theta (t)\right\vert /2$ rather than simply being greater than a constant.
Of special interest is to see whether the time-dependent bound can be
saturated by the use of various types of coherent states in (\ref{GHU}).

Our manuscript is organized as follows: In section 2 we construct the
time-dependent invariant $I(t)$ for the two dimensional harmonic oscillator
on the background described by (\ref{space}). We compute its time-dependent
eigenfunctions $\left\vert \phi _{n}\right\rangle $, determine the phase $%
\alpha (t)$ thereafter and hence the eigenstates $\left\vert \psi
_{n}\right\rangle $ of $H(t)$. As all solutions are dependent on the
solutions of the nonlinear Ermakov-Pinney equation we devote section 3 to a
discussion of its solutions. In section 4 we assemble the solutions from
section 2 and 3 to investigate the validity and quality of a generalized
version of Heisenberg's uncertainty relations. Particular focus is placed on
the study of the uncertainty relations when computed with regard to standard
Glauber coherent states, including their squeezed versions and also Gaussian
Klauder coherent states. In section 5 we state our conclusions.

\section{The 2D harmonic oscillator in a time-dependent background}

The main features of models on time-dependent backgrounds can be explained
by considering simple two dimensional models. Therefore we will examine here
as prototype two dimensional model the harmonic oscillator of the form 
\begin{equation}
H(X,Y,P_{x},P_{y})=\frac{1}{2m}\left( P_{x}^{2}+P_{y}^{2}\right) +\frac{%
m\omega ^{2}}{2}(X^{2}+Y^{2}),  \label{H}
\end{equation}%
on the noncommutative space (\ref{space}). From the many possibly
representations, we choose here a Hermitian one obtained from standard
Bopp-shifts in the conventional canonical variables $x$, $y$, $p_{x}$ and $%
p_{y}$, with nonvanishing commutators $\left[ x,p_{x}\right] =\left[ y,p_{y}%
\right] =i\hbar $, as%
\begin{equation}
X=x-\frac{\theta (t)}{2\hbar }p_{y},\quad Y=y+\frac{\theta (t)}{2\hbar }%
p_{x},\quad P_{x}=p_{x}+\frac{\Omega (t)}{2\hbar }y,\quad P_{y}=p_{y}-\frac{%
\Omega (t)}{2\hbar }x.  \label{XY}
\end{equation}%
As anticipated, when converting the Hamiltonian in (\ref{H}) to the standard
variables it becomes explicitly time-dependent%
\begin{equation}
H(t)=\frac{1}{2}a(t)\left( p_{x}^{2}+p_{y}^{2}\right) +\frac{1}{2}b(t)\left(
x^{2}+y^{2}\right) +c(t)\left( p_{x}y-xp_{y}\right)  \label{Ht}
\end{equation}%
with coefficients%
\begin{equation}
a(t)=\frac{1}{m}+\frac{m\omega ^{2}}{4\hbar ^{2}}\theta ^{2}(t),\quad
b(t)=m\omega ^{2}+\frac{\Omega ^{2}(t)}{4m\hbar ^{2}},\quad c(t)=\frac{%
m\omega ^{2}\theta (t)}{2\hbar }+\frac{\Omega (t)}{2\hbar m}.
\end{equation}%
We notice that for $\theta (t)=0$ we can view this Hamiltonian with an
appropriate identification of the remaining functions as describing a
particle with mass $m$ moving in an axially symmetric electromagnetic field,
see section IV in \cite{Lewis69}. It should also be noted that with a
re-definition of the time-dependent coefficient attempts to solve the
eigenvalue problem related to (\ref{Ht}) can be found in the literature \cite%
{twoDHO1,twoDHO2}. Unfortunately the solutions provided are partly incorrect
or not useful for our purposes as we shall be commenting on below in more
detail.

The quantum equations of motion for the canonical variables associated to
the Hamiltonian (\ref{Ht}) are simply%
\begin{eqnarray}
\dot{x} &=&\frac{1}{i\hbar }[x,H]=a(t)p_{x}+c(t)y,\quad ~~~\ \ \ \ \ \ \dot{y%
}=\frac{1}{i\hbar }[y,H]=a(t)p_{y}-c(t)x,  \label{equ1} \\
\dot{p}_{x} &=&\frac{1}{i\hbar }[p_{x},H]=-b(t)x+c(t)p_{y},\qquad \ \dot{p}%
_{y}=\frac{1}{i\hbar }[p_{y},H]=-b(t)y-c(t)p_{x},  \label{equ2}
\end{eqnarray}%
where we adopt the usual convention for the time derivative $\partial _{t}f=:%
\dot{f}$.

\subsection{Construction of time-dependent invariants}

A non-Hermitian invariant is constructed right away, by following the
argumentation already provided in \cite{Lewis69}. Defining the non-canonical
variables%
\begin{equation}
Q:=(x+iy)e^{i\int\nolimits^{t}c(s)ds}\quad ~~\text{and~~\quad }%
P:=(p_{x}+ip_{y})e^{i\int\nolimits^{t}c(s)ds},
\end{equation}%
satisfying $[Q,P]=0$, we find with (\ref{equ1}) and (\ref{equ2}) the same
equations of motion for these variables%
\begin{equation}
\dot{Q}=a(t)P\quad ~~\text{and~~\quad }\dot{P}=-b(t)Q.
\end{equation}%
as for the harmonic oscillator with a time-dependent mass term \cite{Pedrosa}%
. This is all that matters for the identification of a formal invariant $%
\tilde{I}(t)$ in terms of the variables $Q$ and $P$%
\begin{equation}
\tilde{I}(t)=\frac{1}{2}\left[ \frac{\tau }{\sigma ^{2}}Q^{2}+(\sigma P-%
\frac{\dot{\sigma}}{a}Q)^{2}\right] \neq \tilde{I}^{\dagger }(t),  \label{IT}
\end{equation}%
since we may simply take the expression from the literature and adapt the
relevant quantities appropriately. Here $\sigma $ is a new auxiliary
quantity that has to satisfy a nonlinear Ermakov-Pinney (EP) \cite%
{Ermakov,Pinney} equations including a dissipative term 
\begin{equation}
\ddot{\sigma}-\frac{\dot{a}}{a}\dot{\sigma}+ab\sigma =\tau \frac{a^{2}}{%
\sigma ^{3}},  \label{Erma}
\end{equation}%
with integration constant $\tau $. It is well-known that variations of this
equation are ubiquitous in this context of solving time-dependent
Hamiltonian systems, see for instance equation (5) in \cite{ChoiK}, which
reduces exactly to (\ref{Erma}) for $A\rightarrow a$, $B\rightarrow 0$ and $%
C\rightarrow \tau $ and \cite{Hone,Hawkins,ChoiK2,FBM} for variations of
this equation.

In principle the fact that $\tilde{I}$ in (\ref{IT}) is an invariant means $%
\tilde{I}\tilde{I}^{\dagger }$ or $\tilde{I}^{\dagger }\tilde{I}$ constitute
Hermitian invariants. However, since they will be quartic in the canonical
variables and not directly suitable to an operator approach to find the
corresponding eigensystems we seek an additional one of lower order in the
canonical variables, having however equation (\ref{Erma}) in common.

The symmetry of the Hamiltonian suggest to carry out a quantum canonical
transformation using polar coordinates $x=r\cos \theta $, $y=r\sin \theta $,
which indeed turns out to be very suitable. The canonical coordinates and
momenta are then $r=\sqrt{x^{2}+y^{2}}$, $\theta =\arctan (y/x)$ and $%
p_{r}=\left( xp_{x}+yp_{y}\right) /r-i\hbar /(2r)$, $p_{\theta
}=xp_{y}-yp_{x}$, such that the canonical commutation relations are $%
[r,p_{r}]=[\theta ,p_{\theta }]=i\hbar $. The last term in $p_{r}$ is not
essential for the canonical commutation relations, but its inclusion ensures
the Hermiticity of $p_{r}$ and leads to the convenient identity $%
p_{x}^{2}+p_{y}^{2}=p_{r}^{2}+p_{\theta }^{2}/r^{2}-\hbar ^{2}/(4r^{2})$
allowing to convert the Hamiltonian (\ref{Ht}) into the form%
\begin{equation}
H(t)=\frac{1}{2}a(t)\left( p_{r}^{2}+\frac{p_{\theta }^{2}}{r^{2}}-\frac{%
\hbar ^{2}}{4r^{2}}\right) +\frac{1}{2}b(t)r^{2}-c(t)p_{\theta }.  \label{Hp}
\end{equation}

Applying now the Lewis-Riesenfeld method of invariants and construct a
Hermitian time-dependent invariant $I(t)$ by using (\ref{Inv}), we commence
with the standard assumption that the invariant is of the same order and
form in the canonical variables as the original Hamiltonian. Similarly as
the Hamiltonian, we assume here that also the invariant does not depend
explicitly on $\theta $ and take it to be of the general form%
\begin{equation}
I(t)=\alpha (t)p_{r}^{2}+\beta (t)r^{2}+\gamma (t)\{r,p_{r}\}+\delta (t)%
\frac{p_{\theta }^{2}}{r^{2}}+\varepsilon (t)\frac{p_{\theta }}{r^{2}}+\phi
(t)\frac{1}{r^{2}},  \label{II}
\end{equation}%
with unknown time-dependent coefficients $\alpha (t)$, $\beta (t)$, $\gamma
(t)$ etc. The substitution of (\ref{II}) into (\ref{Inv}) then yields the
following constraints on these coefficients%
\begin{equation}
\dot{\alpha}=-2a\gamma ,\quad \dot{\beta}=2b\gamma ,\quad \dot{\gamma}%
=b\alpha -a\beta ,  \label{3}
\end{equation}%
\begin{equation}
\dot{\delta}p_{\theta }^{2}+\dot{\varepsilon}p_{\theta }+\dot{\phi}=\hbar
^{2}a\gamma -2a\gamma p_{\theta }^{2},\qquad \left( \delta -\alpha \right)
p_{\theta }^{2}+\varepsilon p_{\theta }+\phi +\frac{\alpha \hbar ^{2}}{4}=0.
\label{22}
\end{equation}%
We observe that the equations in (\ref{3}) take on the same form as the
equations underlying the explicit construction for the time-dependent
harmonic oscillator \cite{Pedrosa}. They can be solved by parameterizing $%
\alpha (t)=\sigma ^{2}(t)$ and after one integration we are led exactly to
the nonlinear Ermakov-Pinney equations (\ref{Erma}) underlying the solution
for our non-Hermitian invariant $\tilde{I}(t)$. The remaining equations (\ref%
{22}) are consistently solved by 
\begin{equation}
\delta =\alpha ,\qquad \varepsilon =0,\quad \text{and\quad }\phi =-\frac{%
\alpha \hbar ^{2}}{4}.
\end{equation}%
Assembling everything, the Hermitian invariant $I(t)$ for the time-dependent
Hamiltonian (\ref{Ht}) then acquires the form%
\begin{equation}
I(t)=\frac{\tau }{\sigma ^{2}}r^{2}+\left( \sigma p_{r}-\frac{\dot{\sigma}}{a%
}r\right) ^{2}+\frac{\sigma ^{2}p_{\theta }^{2}}{r^{2}}-\frac{\sigma
^{2}\hbar ^{2}}{4r^{2}},
\end{equation}%
with $\sigma (t)$ determined by the Ermakov-Pinney equation (\ref{Erma}). As
argued already in \cite{Lewis69} the arbitrary constant $\tau $ may be
scaled away, thus that from now on we simply set it to $1$ for convenience
without introducing a new quantity.

Next we solve the eigenvalue equation (\ref{eigen}) by expressing the
invariant $I(t)$ in terms of time-dependent creation and annihilation
operators%
\begin{eqnarray}
\hat{a}(t) &=&\frac{1}{2\sqrt{\hbar }}\left[ \left( \sigma p_{r}-\frac{\dot{%
\sigma}}{a}r\right) -i\left( \frac{r}{\sigma }+\frac{\sigma }{r}(p_{\theta }+%
\frac{\hbar }{2})\right) \right] e^{-i\theta },  \label{a} \\
\qquad \hat{a}^{\dagger }(t) &=&\frac{1}{2\sqrt{\hbar }}e^{i\theta }\left[
\left( \sigma p_{r}-\frac{\dot{\sigma}}{a}r\right) +i\left( \frac{r}{\sigma }%
+\frac{\sigma }{r}(p_{\theta }+\frac{\hbar }{2})\right) \right] ,  \label{ad}
\end{eqnarray}%
satisfying $[\hat{a},\hat{a}^{\dagger }]=1$, by means of the identity 
\begin{equation}
\hbar \left( \hat{a}^{\dagger }\hat{a}+\frac{1}{2}\right) -p_{\theta }=\frac{%
1}{4}I(t)-\frac{1}{2}p_{\theta }=:\hat{I}(t).  \label{Ia}
\end{equation}%
Clearly $\hat{I}(t)$ is also an invariant, where the factor $1/4$ simply
amounts to a new value for the integration constant $\tau $ and $p_{\theta }$
may be added to $I(t)$ since $[H(t),p_{\theta }]=0$.

\subsection{Eigensystem for the time-dependent invariant}

We can now employ the standard argumentation from \cite{Lewis69} to
construct the eigenstates and eigenfunctions for the invariant $\hat{I}(t)$.
Noting first that $[\hat{I}(t),p_{\theta }]=0$, one concludes that $\hat{I}%
(t)$ and $p_{\theta }$ possess simultaneous eigenvectors, say $\left\vert
n,\ell \right\rangle $, with%
\begin{equation}
\hat{I}\left\vert n,\ell \right\rangle =\hbar \left( n+\frac{1}{2}\right)
\left\vert n,\ell \right\rangle ,\qquad p_{\theta }\left\vert n,\ell
\right\rangle =\hbar \ell \left\vert n,\ell \right\rangle ,\qquad
\left\langle n,\ell \right. \left\vert n,\ell \right\rangle =1.
\end{equation}%
Computing therefore $\left\langle n,\ell \right\vert \hat{a}^{\dagger }\hat{a%
}\left\vert n,\ell \right\rangle =n+\ell \geq 0$ implies that for given $n$
we have $\ell \in \{-n,\ldots ,0,1,2,\ldots \}$. The eigenstates of this
sequence therefore obey 
\begin{equation}
\hat{a}\left\vert n,-n\right\rangle =0,\qquad \left\vert n,m-n\right\rangle =%
\frac{1}{\sqrt{m!}}\left( \hat{a}^{\dagger }\right) ^{m}\left\vert
n,-n\right\rangle ,\quad \ \ \text{with }n,m\in \mathbb{N}_{0}.  \label{ann}
\end{equation}%
For all observables that can be expressed in terms of the time-dependent
creation and annihilation operators $\hat{a}^{\dagger }$ and $\hat{a}$, we
can simply use operator techniques to compute their expectation values.
However, the former is not possible for our observables $X$, $Y$, $P_{x}$
and $P_{y}$. We therefore use the explicit representations in coordinate
space $p_{\theta }=-i\hbar \partial _{\theta }$ and $p_{r}=-i\hbar \lbrack
\partial _{r}+1/(2r)]$ to compute the eigenstates. Assuming now $%
\left\langle r,\theta \right. \left\vert n,\ell \right\rangle =$ $\psi
_{n,\ell }(r,\theta )=\varphi _{n}(r)e^{i\ell \theta }$ we have the desired
property $p_{\theta }\psi _{n,\ell }(r,\theta )=\hbar \ell \psi _{n,\ell
}(r,\theta )$. For given $n$, the lowest states are then found from solving
the differential equation $\hat{a}\psi _{n,-n}(r,\theta )=0$, that is 
\begin{equation}
\frac{ie^{-i\theta -i\theta n}}{2ar\sigma \sqrt{\hbar }}\left[ \left( a\hbar
n\sigma ^{2}-ar^{2}+ir^{2}\sigma \text{$\dot{\sigma}$}\right) \varphi
(r)-a\hbar r\sigma ^{2}\partial _{r}\varphi (r)\right] =0.  \label{zero}
\end{equation}%
The solution to (\ref{zero}) is then easily found to be%
\begin{equation}
\psi _{n,-n}(r,\theta )=\lambda _{n}r^{n}e^{-\frac{r^{2}(a-i\sigma \text{$%
\dot{\sigma}$})}{2a\hbar \sigma ^{2}}}e^{-i\theta n},\quad ~~~~\lambda
_{n}^{2}=\frac{1}{\pi n!(\hbar \sigma ^{2})^{(1+n)}}.
\end{equation}%
We have fixed here the constant of integration by demanding the ground state
to be normalized. Subsequently we construct the normalized excited states
from the second relation in (\ref{ann}) to 
\begin{equation}
\psi _{n,m-n}(r,\theta )=\lambda _{n}\frac{\left( i\hbar ^{1/2}\sigma
\right) ^{m}}{\sqrt{m!}}r^{n-m}e^{i\theta (m-n)-\frac{a-i\sigma \text{$\dot{%
\sigma}$}}{2a\hbar \sigma ^{2}}r^{2}}U\left( -m,1-m+n,\frac{r^{2}}{\hbar
\sigma ^{2}}\right) ,
\end{equation}%
with $U(a,b,z)$ denoting the confluent hypergeometric function. The
orthonormality relation $\int\nolimits_{0}^{2\pi }d\theta
\int\nolimits_{0}^{\infty }dr~r\psi _{n,m-n}^{\ast }(r,\theta )\psi
_{n^{\prime },m^{\prime }-n^{\prime }}(r,\theta )=\delta _{nn^{\prime
}}\delta _{mm^{\prime }}$ is verified by using the standard properties of
the latter function.

It should be noted here that our solution differs from those found in the
literature \cite{twoDHO1,twoDHO2}. As was pointed out in \cite{twoDHO2} the
solutions provided in \cite{twoDHO1} are incorrect as they lead to
time-dependent eigenvalues and thus contradict the basic foundations of the
Lewis-Riesenfeld theory, i.e. equation (\ref{eigen}). Our solution differs
also slightly from those in \cite{twoDHO2}. Moreover, in \cite{twoDHO2} the
normalization constant was left undetermined, which is, however, crucial in
concrete computations following below.

\subsection{Eigensystem for the Hamiltonian}

The last step in the Lewis-Riesenfeld procedure consists of computing the
phase $\alpha (t)$ in (\ref{ph}) by solving the equation%
\begin{equation}
\dot{\alpha}_{n,\ell }=\frac{1}{\hbar }\left\langle n,\ell \right\vert
i\hbar \partial _{t}-H\left\vert n,\ell \right\rangle .  \label{alpha}
\end{equation}%
As already argued in \cite{Lewis69}, this may be achieved by constructing a
recursive equation for the right hand side of (\ref{alpha}), computing some
explicit expectation values, using the freedom to choose the phase for the
vacuum state and a subsequent integration.

We commence by simply replacing $\left\vert n,\ell \right\rangle =\hat{a}%
^{\dagger }/\sqrt{n+\ell }\left\vert n,\ell -1\right\rangle $ in (\ref{alpha}%
), obtaining%
\begin{equation}
\left\langle n,\ell \right\vert i\hbar \partial _{t}-H\left\vert n,\ell
\right\rangle =\left\langle n,\ell -1\right\vert i\hbar \partial
_{t}-H\left\vert n,\ell -1\right\rangle +\frac{1}{n+\ell }\left\langle
n,\ell -1\right\vert \left[ \hat{a},i\hbar \partial _{t}-H\right] \hat{a}%
^{\dagger }\left\vert n,\ell -1\right\rangle .  \label{rec}
\end{equation}%
Using next the expression (\ref{a}) for the annihilation operator and the
Hamiltonian in polar coordinates (\ref{Hp}), we compute%
\begin{equation}
\left[ \hat{a},i\hbar \partial _{t}-H\right] =\hbar \left( c(t)-\frac{a(t)}{%
\sigma ^{2}(t)}\right) \hat{a},  \label{comm}
\end{equation}%
upon replacing $\ddot{\sigma}$ by means of the EP-equation in the form (\ref%
{Erma}). Substitution of (\ref{comm}) into (\ref{rec}) allows for the
computation of the expectation value, thus leading to the recursive equation%
\begin{equation}
\left\langle n,\ell \right\vert i\hbar \partial _{t}-H\left\vert n,\ell
\right\rangle =\left\langle n,\ell -1\right\vert i\hbar \partial
_{t}-H\left\vert n,\ell -1\right\rangle +\hbar \left( c(t)-\frac{a(t)}{%
\sigma ^{2}(t)}\right) .  \label{rec2}
\end{equation}%
We may now iterate this equation until we reach the expectation values for
vacuum state$~\left\langle n,-n\right\vert i\hbar \partial _{t}-H\left\vert
n,-n\right\rangle $. As argued in \cite{Lewis69}, the matrix element $%
\left\langle n,-n\right\vert \partial _{t}\left\vert n,-n\right\rangle $
involves an arbitrary constant, which we conveniently choose to set to $%
\left\langle n,-n\right\vert \partial _{t}\left\vert n,-n\right\rangle
=\left\langle n,-n\right\vert H\left\vert n,-n\right\rangle $. Therefore we
obtain the expectation value%
\begin{equation}
\left\langle n,\ell \right\vert i\hbar \partial _{t}-H\left\vert n,\ell
\right\rangle =(n+\ell )\hbar \left( c(t)-\frac{a(t)}{\sigma ^{2}(t)}\right)
,
\end{equation}%
allowing us to compute the phase to%
\begin{equation}
\alpha _{n,\ell }(t)=(n+\ell )\int\nolimits^{t}\left( c(s)-\frac{a(s)}{%
\sigma ^{2}(s)}\right) ds.
\end{equation}%
Our result for $\alpha _{n,\ell }(t)$ differs from the phase computed in 
\cite{twoDHO2}, where the $c(s)$-term is absent.

We have now obtained explicit eigenfunctions for the Hamiltonian (\ref{H})
for any time-dependent background field in terms of the solutions of the
EP-equation. Mostly in the literature the analysis is abandoned at this
stage and the invariants and wavefunctions are simply expressed in terms of
the yet to be determined solution to the EP-equation. However, for concrete
computations of measurable quantities one needs to address the auxiliary
problem and solve the equations explicitly for the time-dependent functions
appearing in the Hamiltonian. Surprisingly little attention has been paid to
this problem in the context of solving time-dependent Hamiltonian systems
and therefore we will discuss the solutions of our auxiliary equation (\ref%
{Erma}) in the next subsection.

\section{The Ermakov-Pinney equation}

The simplest special solution arises when taking $\theta (t)=const$, such
that $\dot{a}=0$ and consequently the dissipative term vanishes. For this
case particular solutions were already found by Pinney \cite{Pinney}%
\begin{equation}
\sigma =\sqrt{u_{1}^{2}+\tau a^{2}\frac{u_{2}^{2}}{W^{2}}},  \label{SolEP}
\end{equation}%
where $u_{1}$, $u_{2}$ are the two linearly independent solutions of the
equation%
\begin{equation}
\ddot{u}+ab(t)u=0,  \label{EPl}
\end{equation}%
and $W=u_{1}\dot{u}_{2}-\dot{u}_{1}u_{2}$ is the corresponding Wronskian.

When $\dot{a}\neq 0$ no general solution to (\ref{Erma}) is known, although
one can construct a variety of explicit solutions following the procedure
proposed in \cite{MancasRosu1,MancasRosu2}. We briefly outline the method
and use it to construct some new solutions, which we employ later on. We
start by considering the ordinary differential equation of the general form%
\begin{equation}
\frac{d^{2}\sigma }{dt^{2}}+g(\sigma )\frac{d\sigma }{dt}+h(\sigma )=0,
\label{ODE}
\end{equation}%
for which the EP-equation can be seen as a special case with the appropriate
choices for $g(\sigma )$ and $h(\sigma )$. Introducing the new quantity $%
\eta (\sigma ):=d\sigma /dt$, the equation (\ref{ODE}) is easily converted
into the first order differential equation%
\begin{equation}
\eta \frac{d\eta }{d\sigma }+g(\sigma )\eta +h(\sigma )=0.  \label{ODE2}
\end{equation}%
This implies that when having solved (\ref{ODE2}), a solution to the
original equation (\ref{ODE}) can be obtained simply from inverting $%
\int^{\sigma }\eta ^{-1}(s)ds=t$. It can be shown by direct substitution
that (\ref{ODE2}) admits the solution%
\begin{equation}
\eta (\sigma )=\lambda _{\kappa }\frac{h(\sigma )}{g(\sigma )}\qquad \text{%
with }\lambda _{\kappa }^{\pm }=\frac{-1\pm \sqrt{1-4\kappa }}{2\kappa },
\end{equation}%
if the Chiellini integrability condition \cite{Chiellini} 
\begin{equation}
\frac{d}{d\sigma }\left( \frac{h(\sigma )}{g(\sigma )}\right) =\kappa
g(\sigma ),  \label{CIC}
\end{equation}%
with $\kappa \in \mathbb{R}$ holds. Based on this we may then find exact
analytical solutions for instance by starting with a given $g(\sigma )$ and
subsequently compute 
\begin{equation}
\eta (\sigma )=\kappa \lambda _{\kappa }\int^{\sigma }g(s)ds\qquad \text{%
and\qquad }h(\sigma )=\kappa g(\sigma )\int^{\sigma }g(s)ds,  \label{gh}
\end{equation}%
or by starting with a given $h(\sigma )$ and subsequently evaluate 
\begin{equation}
\eta (\sigma )=\pm \lambda _{\kappa }\sqrt{2\kappa \int^{\sigma }h(s)ds}%
\qquad \text{and\qquad }g(\sigma )=\frac{h(\sigma )}{\sqrt{2\kappa
\int^{\sigma }h(s)ds}}.  \label{hg}
\end{equation}%
Following this solution procedure means of course that we are not
pre-selecting our background fields $\theta (t)$ and $\Omega (t)$, but
instead we determine them by primarily insisting on the integrability of the
EP-equation. The virtue of this method is that it leads to exact solutions.
Nonetheless, one might also be interested in concrete types of background
fields for which the integrability condition (\ref{CIC}) does not hold, in
which case we will resort to a numerical analysis.

\subsection{Non-dissipative solutions}

For the special case $\theta (t)=const$, i.e. $\dot{a}=0$ we can simply
pre-select any explicit form for $\Omega (t)$, and thereby $b(t)$, to
construct the solutions from the general formula (\ref{SolEP}). For instance
for $a(t)=\alpha $ and $b(t)=\beta e^{\gamma t}$, $\alpha ,\beta ,\gamma \in 
\mathbb{R}$, we solve (\ref{EPl}) in terms of Bessel functions and
subsequently obtain the particular solution by means of (\ref{SolEP}) 
\begin{equation}
\sigma (t)=\sqrt{\frac{\pi ^{2}\alpha ^{2}\tau {}}{\gamma ^{2}c_{1}^{2}}%
Y_{0}^{2}\left( \frac{2\sqrt{\alpha \beta }e^{\gamma t/2}}{\gamma }\right)
+c_{1}^{2}J_{0}^{2}\left( \frac{2\sqrt{\alpha \beta }e^{\gamma t/2}}{\gamma }%
\right) {}},
\end{equation}%
with integration constant $c_{1}\in \mathbb{R}$ and $J_{0}$, $Y_{0}$
denoting the Bessel functions of first and second kind, respectively.
Similarly different solutions are easily constructed for any other explicit
choice of $b(t)$ for which (\ref{EPl}) admits a solution.

\subsection{Exponentially decaying solutions}

Let us now switch on the dissipative term and take $\dot{a}\neq 0$ by making
the additional assumption $g(\sigma )=\gamma \in $ $\mathbb{R}$. Then the
second equation in (\ref{gh}) together with the explicit form of $h(\sigma )$
from (\ref{Erma}) yields the consistency equation%
\begin{equation}
\kappa \gamma ^{2}\sigma =ab\sigma -\tau \frac{a^{2}}{\sigma ^{3}},
\label{con}
\end{equation}%
from which we deduce that $ab=const$ and $a\sim \sigma ^{2}$. Since we may
find $a(t)$ simply from $-\dot{a}/a=\gamma $, all other functions follow
from the proportionality relations. We find exponentially decaying and
increasing background fields corresponding to exponentially decaying
solutions of the EP-equation%
\begin{equation}
a(t)=\alpha e^{-\gamma t},\quad b(t)=\beta e^{\gamma t},\quad \text{and\quad 
}\sigma (t)=\mu e^{-\gamma t/2},  \label{sol1}
\end{equation}%
with $\alpha $, $\beta $, $\gamma \in \mathbb{R}$, together with the
constraint $\mu ^{4}=\tau \alpha ^{2}/(\alpha \beta -\kappa \gamma ^{2})$
resulting from (\ref{con}). The Chiellini constant $\kappa $ is not fixed at
this point, but simply determined by substituting the expressions from (\ref%
{sol1}) into (\ref{Erma}), leading to $\kappa =1/4$. A special case of our
solution corresponds to the one reported in \cite{ChoiK} where the
EP-equation of the type (\ref{Erma}) appears as an auxiliary equation in the
solution procedure for the Caldirola-Kanai Hamiltonian \cite{Caldirola,Kanai}%
.

Notice that for our background fields the requirement that $\theta
(t),\Omega (t)\in \mathbb{R}$ implies that this solution leads to cutoff
times $t_{c}$ after which the background field needs to be vanishing, that
is $t<t_{c}=\ln (m\alpha )/\gamma $ for $\alpha ,\gamma >0$. It should also
be noted that the constraint on the constants is quite severe and one might
change the overall qualitative behaviour of the solution from a decaying
solution to an oscillatory behaviour when relaxing the integrability
condition.

\begin{figure}[h]
\centering   \includegraphics[width=7.5cm,height=6.0cm]{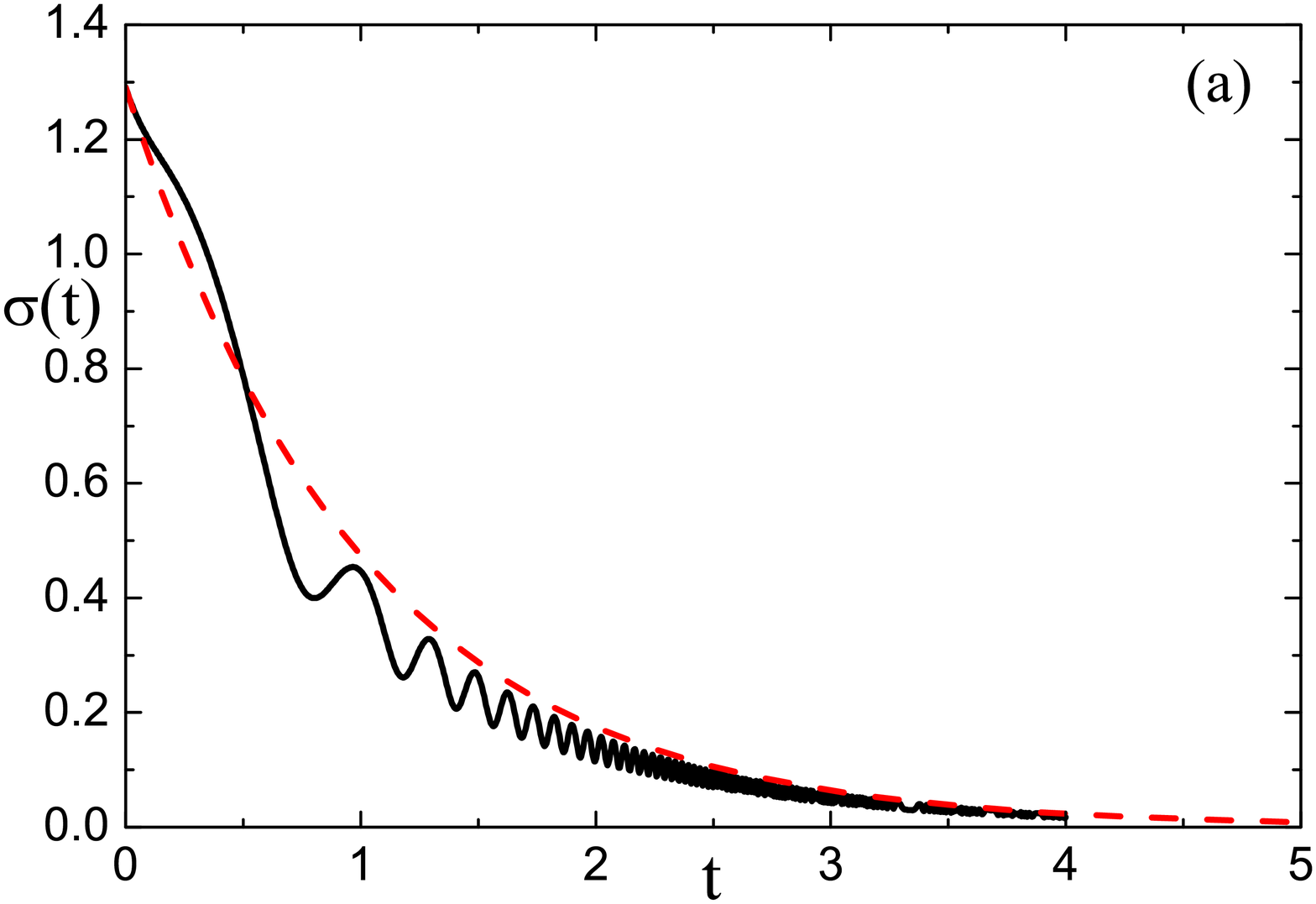} %
\includegraphics[width=7.5cm,height=6.0cm]{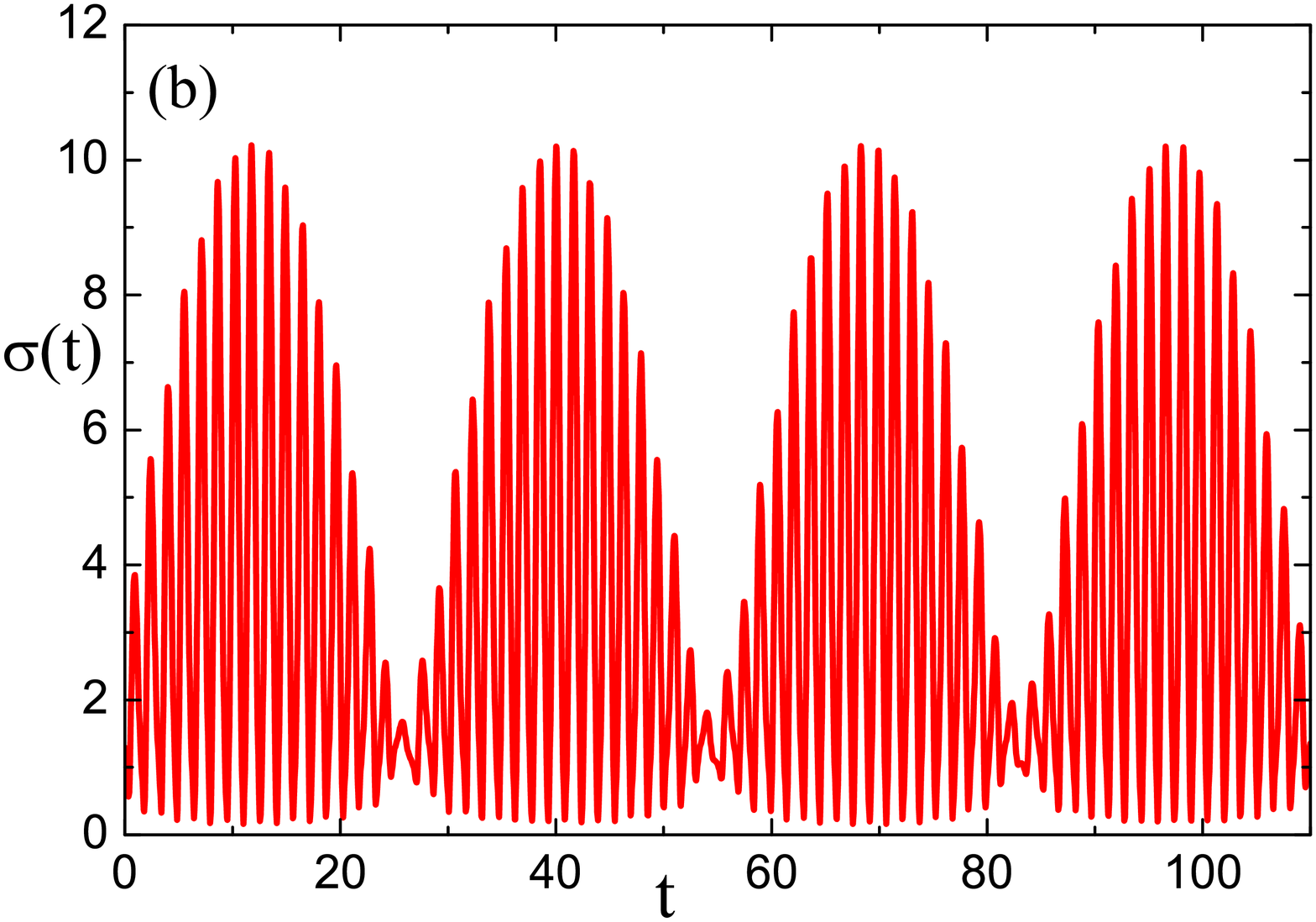} \centering   
\caption{(a) Exactly integrable solution (\protect\ref{sol1}) (red, dashed)
versus a non-Chiellini integrable solution for pre-selected exponential
backgrounds $\protect\theta (t)=\protect\alpha e^{-\protect\gamma t}$ and $%
\Omega (t)=\protect\beta e^{\protect\gamma t}$ (black, solid). (b)
Non-Chiellini integrable solution for pre-selected sinusoidal background $%
\protect\theta (t)=\protect\alpha \sin (\protect\gamma t)$ and $\Omega (t)=%
\protect\beta \sin (\protect\gamma t/2)$. In both panels the constants are $%
\protect\alpha=5$, $\protect\beta=2$, $\protect\gamma=2$, $m=\hbar=\protect%
\tau=\protect\omega=1$, $\protect\kappa=1/4$ and $\protect\mu=\protect\sqrt{%
5/3}$.}
\label{fig1}
\end{figure}

\subsection{Rationally decaying solutions}

Next we assume $g(\sigma )=\gamma \sigma ^{n}$ with $n\in \mathbb{N}$. The
consistency equation then reads%
\begin{equation}
\kappa \gamma ^{2}\frac{\sigma ^{2n+1}}{n+1}=ab\sigma -\tau \frac{a^{2}}{%
\sigma ^{3}},
\end{equation}%
which implies that $ab\sim \sigma ^{2n}$ and $a\sim \sigma ^{n+2}$.
Determining $a(t)$ simply from $-\dot{a}/a=\gamma \sigma ^{n}$, we compute
all other functions from the proportionality relations. We find rational
solutions to the background fields and the EP-equation%
\begin{equation}
a(t)=\frac{\alpha \left( \frac{n+2}{n}\right) ^{\frac{n+2}{n}}}{(\gamma
t-\mu )^{(n+2)/n}},\quad b(t)=\frac{\beta \left( \frac{n}{n+2}\right) ^{%
\frac{2}{n}-1}}{(\gamma t-\mu )^{1-\frac{2}{n}}},\quad \text{and\quad }%
\sigma (t)=\frac{\left( \frac{n+2}{n}\right) ^{\frac{1}{n}}}{(\gamma t-\mu
)^{1/n}},
\end{equation}%
with constraint $\gamma ^{2}=(n+1)(\alpha \beta -\tau \alpha ^{2})/\kappa $.
The Chiellini constant is subsequently fixed to $\kappa =(n+1)/(n+2)^{2}$.\
To maintain real solutions requires here a cutoff time $t<t_{c}=\mu /\gamma $
for $\gamma ,\mu >0$.

\subsection{Non-Chiellini integrable solutions with pre-selected background}

As pointed out, the solutions constrcuted in the previous subsections are
special in the sense that the Chiellini integrability has been superimposed
onto them. Nonetheless, given a specific background we may always find
numerical solutions. In figure \ref{fig1} we depict some solutions for
exponential and sinusoidal background fields which we shall employ below in
our solutions for the time-dependent wavefunctions.

\section{The generalized uncertainty relations}

\subsection{The generalized uncertainty relations for eigenstates}

We have assembled now all the necessary ingredients for the explicit
computation of expectation values. We are therefore in the position to test
the generalized uncertainty relations (\ref{GHU}). Having obtained explicit
expressions for the wavefunctions in coordinate space, we simply use the
representation in polar coordinates $x=r\cos \theta $, $y=r\sin \theta $, $%
p_{x}=-i\hbar \cos \theta \partial _{r}+i\hbar /r\sin \theta \partial
_{\theta }$, $p_{y}=-i\hbar \sin \theta \partial _{r}-i\hbar /r\cos \theta
\partial _{\theta }$ and the corresponding relations for the operators in (%
\ref{XY}) to compute the relevant matrix elements. We comence with the
verification of the standard uncertainty relations for the auxiliary
variables $x,y,$ $p_{x},$ $p_{y}$. By evaluating the explicit integrals we
obtain their matrix elements%
\begin{eqnarray}
\left\langle n,m-n\right\vert x\left\vert n,m^{\prime }-n\right\rangle  &=&i%
\frac{\sqrt{\hbar }}{2}\sigma \left( \sqrt{m^{\prime }}e^{i\alpha
_{0,1}}\delta _{m^{\prime },m+1}-\sqrt{m}e^{-i\alpha _{0,1}}\delta
_{m,m^{\prime }+1}\right) ,  \label{MA1} \\
\left\langle n,m-n\right\vert y\left\vert n,m^{\prime }-n\right\rangle  &=&-%
\frac{\sqrt{\hbar }}{2}\sigma \left( \sqrt{m^{\prime }}e^{i\alpha
_{0,1}}\delta _{m^{\prime },m+1}+\sqrt{m}e^{-i\alpha _{0,1}}\delta
_{m,m^{\prime }+1}\right) ,  \label{MA2} \\
\left\langle n,m-n\right\vert p_{x}\left\vert n,m^{\prime }-n\right\rangle 
&=&\frac{\sqrt{\hbar }}{2}\left[ \chi _{+}\sqrt{m^{\prime }}e^{i\alpha
_{0,1}}\delta _{m^{\prime },m+1}+\chi _{-}\sqrt{m}e^{-i\alpha _{0,1}}\delta
_{m,m^{\prime }+1}\right] ,  \label{MA3} \\
\left\langle n,m-n\right\vert p_{y}\left\vert n,m^{\prime }-n\right\rangle 
&=&i\frac{\sqrt{\hbar }}{2}\left[ \chi _{+}\sqrt{m^{\prime }}e^{i\alpha
_{0,1}}\delta _{m^{\prime },m+1}-\chi _{-}\sqrt{m}e^{-i\alpha _{0,1}}\delta
_{m,m^{\prime }+1}\right] ,  \label{MA4}
\end{eqnarray}%
and%
\begin{eqnarray}
\!\!\!\left\langle n,m-n\right\vert x^{2},y^{2}\left\vert n,m^{\prime
}-n\right\rangle  &=&\frac{\hbar }{2}(n+m+1)\sigma ^{2}\delta _{m,m^{\prime
}}\mp \frac{\hbar \sigma ^{2}}{2\sqrt{2}}\mu (m,m^{\prime })e^{i\alpha
_{0,2}}\delta _{m^{\prime },m+2}  \notag \\
&&\mp \frac{\hbar \sigma ^{2}}{2\sqrt{2}}\mu (m^{\prime },m)e^{-i\alpha
_{0,2}}\delta _{m,m^{\prime }+2},  \label{ma1} \\
\!\!\!\left\langle n,m-n\right\vert p_{x}^{2},p_{y}^{2}\left\vert
n,m^{\prime }-n\right\rangle  &=&\frac{\hbar }{2}(n+m+1)\chi _{+}\chi
_{-}\delta _{m,m^{\prime }}\pm \frac{\hbar \chi _{+}^{2}}{2\sqrt{2}}\mu
(m,m^{\prime })e^{i\alpha _{0,2}}\delta _{m^{\prime },m+2}  \notag \\
&&\pm \frac{\hbar \chi _{-}^{2}}{2\sqrt{2}}\mu (m^{\prime },m)e^{-i\alpha
_{0,2}}\delta _{m,m^{\prime }+2},  \label{ma2} \\
\left\langle n,m-n\right\vert xp_{y}\left\vert n,m^{\prime }-n\right\rangle
&=&\frac{\hbar }{2}(m-n)\delta _{m,m^{\prime }}-\frac{\hbar \sigma \chi _{+}%
}{2\sqrt{2}}\mu (m,m^{\prime })e^{i\alpha _{0,2}}\delta _{m^{\prime },m+2} 
\notag \\
&&-\frac{\hbar \sigma \chi _{-}}{2\sqrt{2}}\mu (m^{\prime },m)e^{-i\alpha
_{0,2}}\delta _{m,m^{\prime }+2},  \label{ma3} 
\end{eqnarray}

\begin{eqnarray}
\left\langle n,m-n\right\vert yp_{x}\left\vert n,m^{\prime }-n\right\rangle 
&=&\frac{\hbar }{2}(n-m)\delta _{m,m^{\prime }}-\frac{\hbar \sigma \chi _{+}%
}{2\sqrt{2}}\mu (m,m^{\prime })e^{i\alpha _{0,2}}\delta _{m^{\prime },m+2} 
\notag \\
&&-\frac{\hbar \sigma \chi _{-}}{2\sqrt{2}}\mu (m^{\prime },m)e^{-i\alpha
_{0,2}}\delta _{m,m^{\prime }+2},  \label{ma4}
\end{eqnarray}%
where we abbreviated $\chi _{\pm }:=\frac{1}{\sigma }\pm i\frac{\dot{\sigma}%
}{a}$ and $\mu (x,y):=\sqrt{\left( \frac{x}{2}+1\right) (y-1)}$. 

Using the above expressions the relevant variances are computed to%
\begin{eqnarray}
\left. \Delta x\right\vert _{\psi _{n,m-n}}^{2} &=&\left. \Delta
y\right\vert _{\psi _{n,m-n}}^{2}=\frac{\hbar }{2}(n+m+1)\sigma ^{2}, \\
\left. \Delta p_{x}\right\vert _{\psi _{n,m-n}}^{2} &=&\left. \Delta
p_{y}\right\vert _{\psi _{n,m-n}}^{2}=\frac{\hbar }{2}(n+m+1)\left( \frac{1}{%
\sigma ^{2}}+\frac{\dot{\sigma}^{2}}{a^{2}}\right) .
\end{eqnarray}%
It is then easy to verify that the standard uncertainty relations indeed
hold 
\begin{eqnarray}
\left. \Delta x\Delta p_{x}\right\vert _{\psi _{n,m-n}} &=&\left. \Delta
y\Delta p_{y}\right\vert _{\psi _{n,m-n}}=\frac{\hbar }{2}(n+m+1)\sqrt{1+%
\frac{\sigma ^{2}\dot{\sigma}^{2}}{a^{2}}}\geq \frac{\hbar }{2}, \\
\left. \Delta x\Delta y\right\vert _{\psi _{n,m-n}} &=&\frac{\hbar }{2}%
(n+m+1)\sigma ^{2}\geq 0, \\
\left. \Delta p_{x}\Delta p_{y}\right\vert _{\psi _{n,m-n}} &=&\frac{\hbar }{%
2}(n+m+1)\left( \frac{1}{\sigma ^{2}}+\frac{\dot{\sigma}^{2}}{a^{2}}\right)
\geq 0.
\end{eqnarray}%
However, for our model (\ref{H}) these quantities are mere auxiliary
objects. Therefore, we need to compute the corresponding relations for the
noncommutative quantities in our original system (\ref{H}) on the
time-dependent background. In the light of (\ref{space}) and (\ref{GHU})
they should produce a generalized version of the uncertainty relations with
a time-dependent lower bound. We find $\left\langle n,m-n\right\vert 
\mathcal{O}\left\vert n,m-n\right\rangle =0$ for $\mathcal{O}=X,Y,$ $P_{x},$ 
$P_{y}$, not reported here, and afterwards%
\begin{eqnarray}
\left. \Delta X\right\vert _{\psi _{n,m-n}}^{2} &=&\left. \Delta
Y\right\vert _{\psi _{n,m-n}}^{2}=\left. \Delta x\right\vert _{\psi
_{n,m-n}}^{2}+\frac{n-m}{2}\theta (t)+\frac{n+m+1}{8\hbar }\left( \frac{1}{%
\sigma ^{2}}+\frac{\dot{\sigma}^{2}}{a^{2}}\right) \theta ^{2}(t),~~~~~~~ \\
\left. \Delta P_{x}\right\vert _{\psi _{n,m-n}}^{2} &=&\left. \Delta
P_{y}\right\vert _{\psi _{n,m-n}}^{2}=\left. \Delta p_{x}\right\vert _{\psi
_{n,m-n}}^{2}+\frac{n-m}{2}\Omega (t)+\frac{n+m+1}{8\hbar }\sigma ^{2}\Omega
^{2}(t),
\end{eqnarray}%
from which we deduce the generalized version of the uncertainty relations%
\begin{eqnarray}
\left. \Delta X\Delta Y\right\vert _{\psi _{n,m-n}} &=&\frac{n-m}{2}\theta
(t)+\frac{n+m+1}{8\hbar }\left[ 4\hbar \sigma ^{2}+\left( \frac{1}{\sigma
^{2}}+\frac{\dot{\sigma}^{2}}{a^{2}}\right) \theta ^{2}(t)\right] \geq \frac{%
\theta (t)}{2},  \label{in1} \\
\left. \Delta P_{x}\Delta P_{y}\right\vert _{\psi _{n,m-n}} &=&\frac{\hbar }{%
2}(n+m+1)\left[ \frac{\sigma ^{2}\Omega ^{2}(t)}{4}+\left( \frac{1}{\sigma
^{2}}+\frac{\dot{\sigma}^{2}}{a^{2}}\right) \right] +\frac{n-m}{2}\Omega
(t)\geq \frac{\Omega (t)}{2},~~~~~  \label{in2} \\
\left. \Delta X\Delta P_{x}\right\vert _{\psi _{n,m-n}} &=&\left. \Delta
Y\Delta P_{y}\right\vert _{\psi _{n,m-n}}\geq \frac{\hbar }{2}+\frac{\theta
(t)\Omega (t)}{8\hbar }.  \label{in3}
\end{eqnarray}%
To prove the validity of these inequalities we note for instance that the
smallest value for the left hand side of (\ref{in1}) results from $\left.
\Delta X\Delta Y\right\vert _{\psi _{0,0}}$. Therefore demonstrating that
the quantity $f[\theta (t)]:=\left. \Delta X\Delta Y\right\vert _{\psi
_{0,0}}-\theta (t)/2$ is always nonnegative will establish (\ref{in1}).
Noting for this purpose that $f[0]=$ $\hbar \sigma ^{2}/2$, $\lim_{\theta
(t)\rightarrow \infty }f[\theta (t)]\rightarrow \infty $ and that the local
minimum at $\theta _{\min }(t)=2\hbar \sigma ^{2}a^{2}/(a^{2}+\sigma ^{2}%
\dot{\sigma}^{2})$ acquires the value $f[\theta _{\min }(t)]=\hbar \sigma
^{4}\dot{\sigma}^{2}/(2a^{2}+2\sigma ^{2}\dot{\sigma}^{2})\geq 0$ guarantees
that $f[\theta (t)]\geq 0$ and therefore the validity of (\ref{in1}). One
may argue similarly for (\ref{in2}) and (\ref{in3}), which we will not
present here.

In order to display the deviation from the lower bound we depict in figure
2-4 the uncertainty for backgrounds corresponding to the solutions of the
EP-equation displayed in figure 1. As expected from our analytical
expressions in (\ref{in1}) and previous results, the smallest uncertainties
are observed for the smaller quantum numbers.

\begin{figure}[h]
\centering   \includegraphics[width=7.5cm,height=6.0cm]{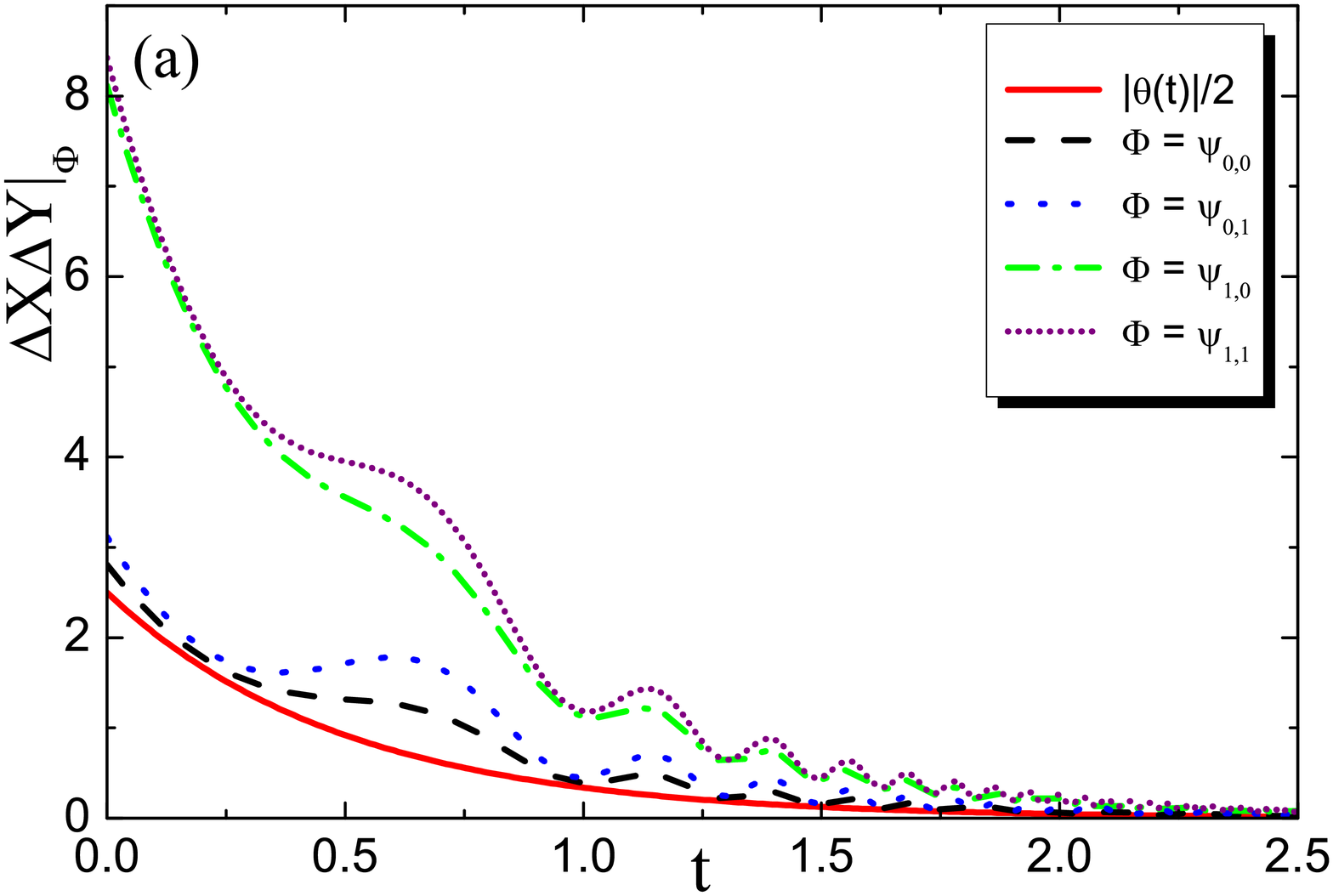} %
\includegraphics[width=7.5cm,height=6.0cm]{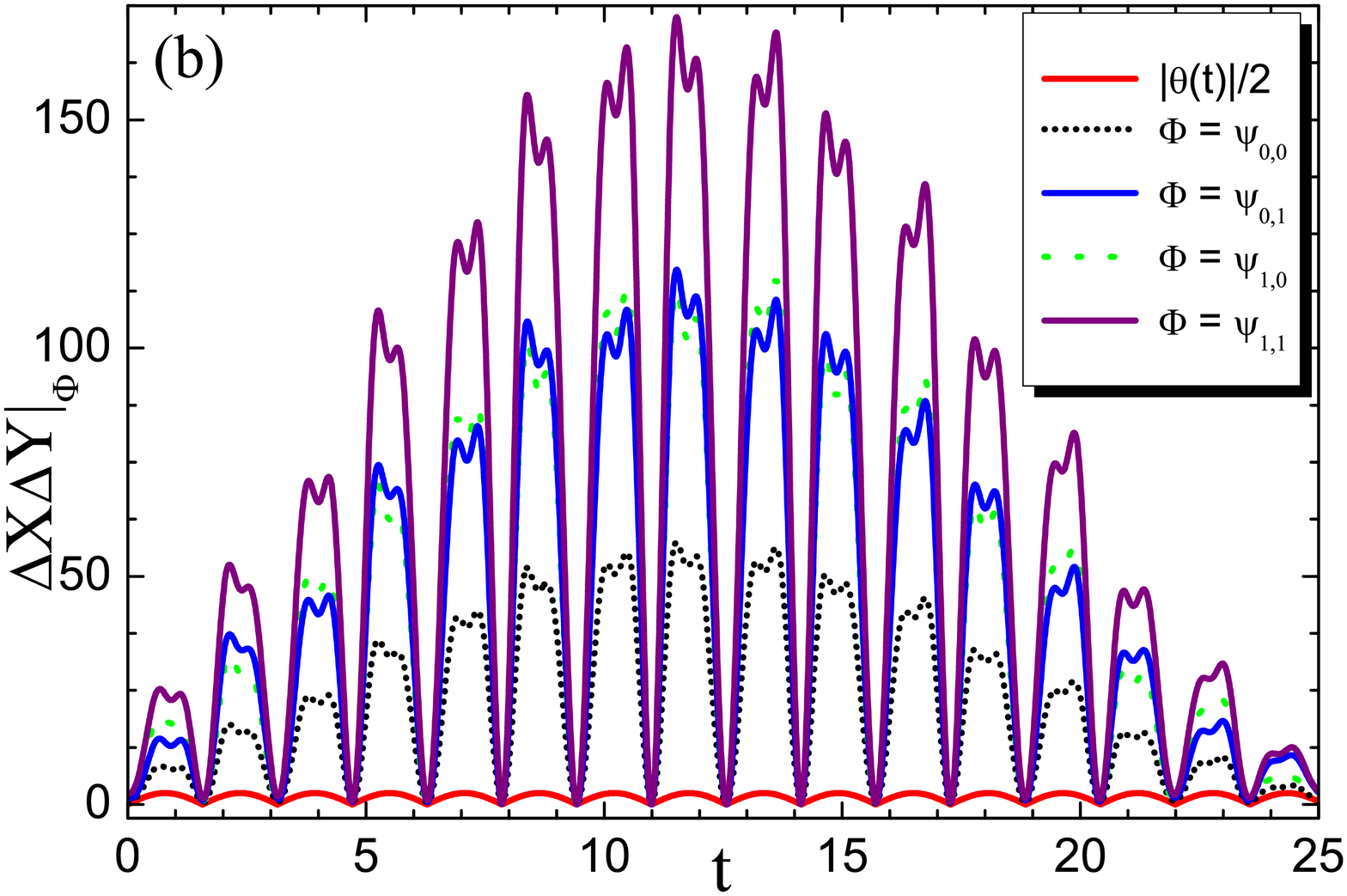} \centering   
\caption{Uncertainties $\left. \Delta X\Delta Y\right\vert _{\protect\psi %
_{n,m-n}}$ versus the generalized lower bound (a) for background fields $%
\protect\theta (t)=\protect\alpha e^{-\protect\gamma t}$ and $\Omega (t)=%
\protect\beta e^{\protect\gamma t}$ and (b) for background fields $\protect%
\theta (t)=\protect\alpha \sin (\protect\gamma t)$ and $\Omega (t)=\protect%
\beta \sin (\protect\gamma t/2)$. In both panels the constants are $\protect%
\alpha=5$, $\protect\beta=2$, $\protect\gamma=2$, $m=\hbar=\protect\tau=%
\protect\omega=1$, $\protect\kappa=1/4$ and $\protect\mu=\protect\sqrt{5/3}$%
. }
\label{fig2}
\end{figure}

\begin{figure}[h]
\centering   \includegraphics[width=7.5cm,height=6.0cm]{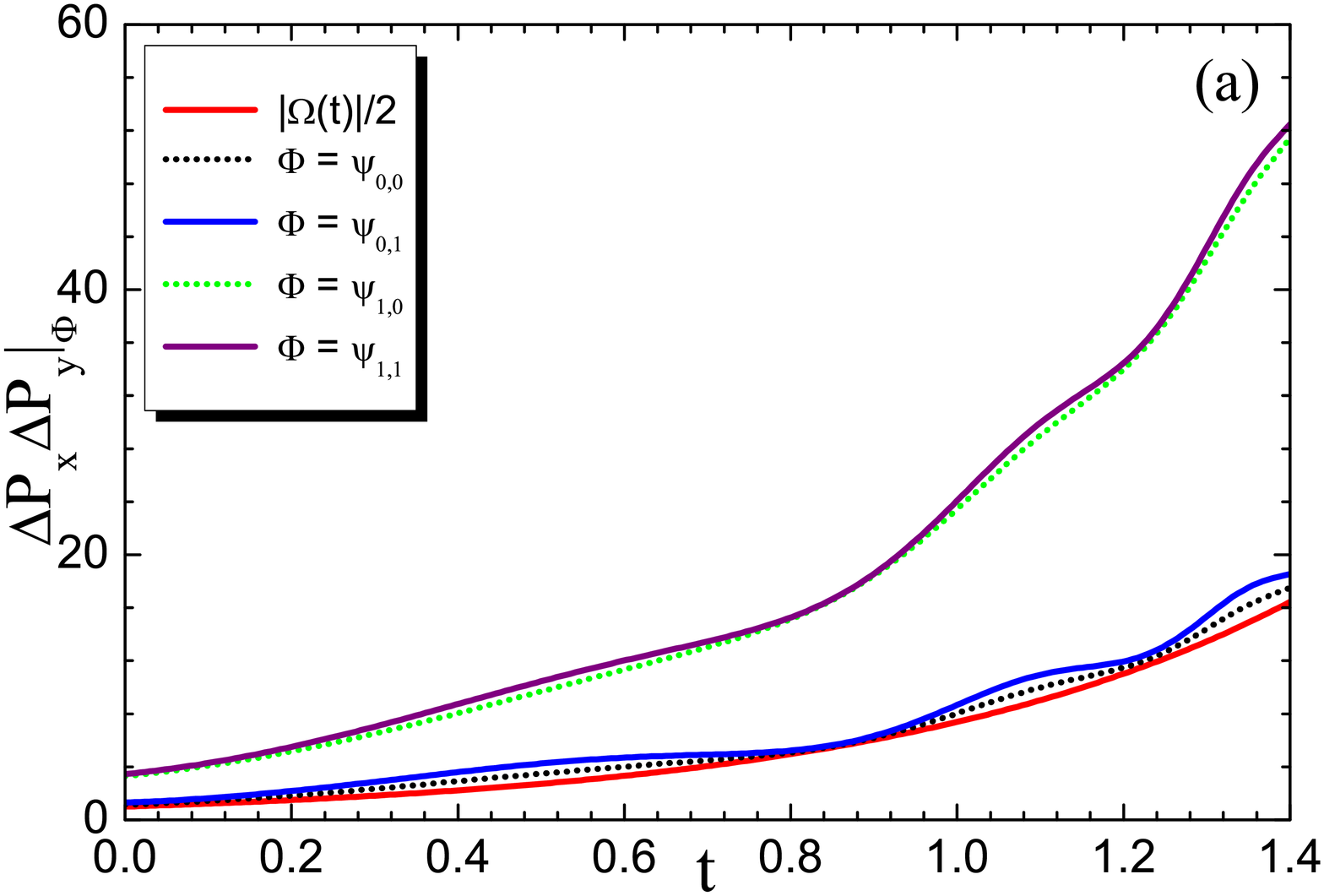} %
\includegraphics[width=7.5cm,height=6.0cm]{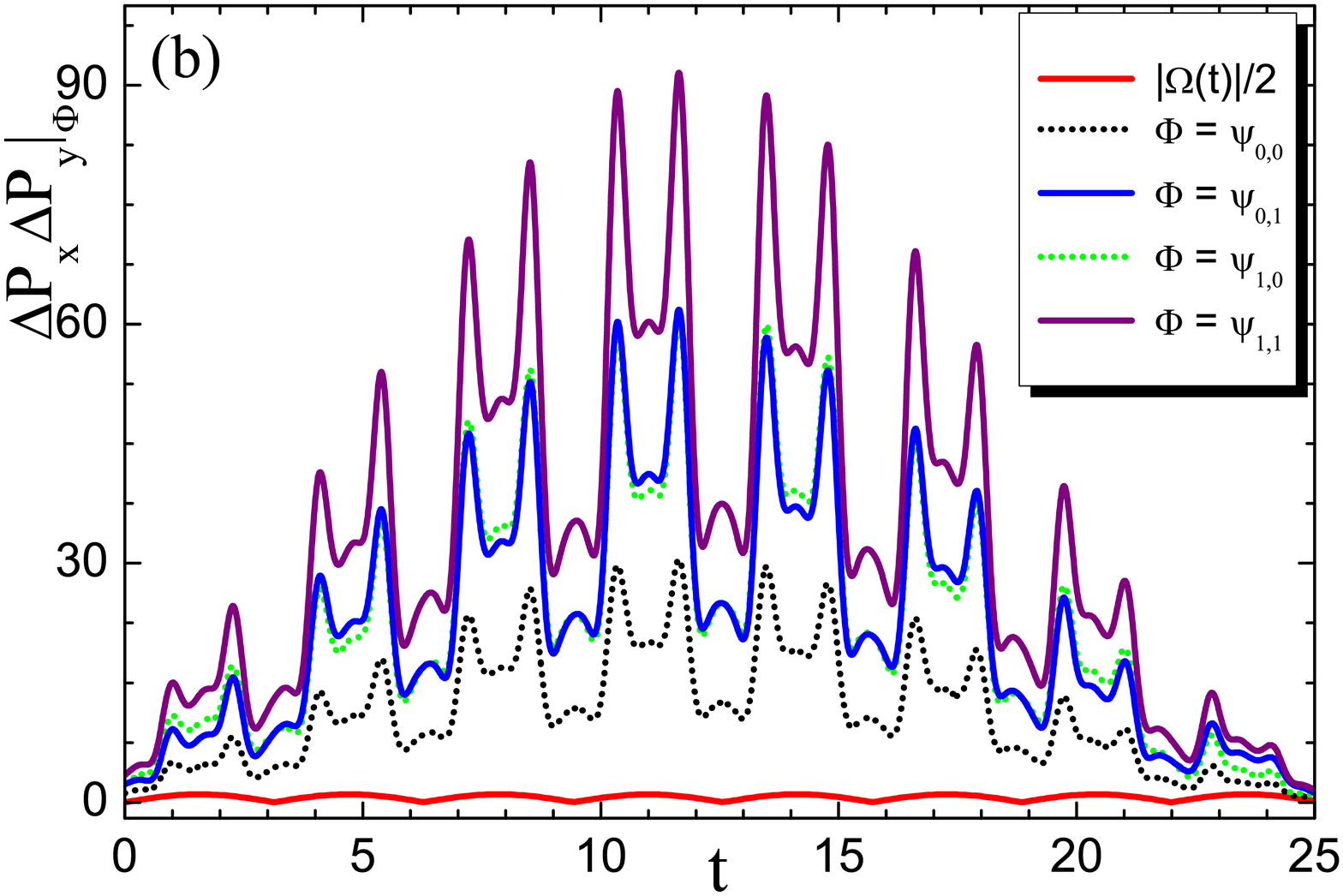} \centering   
\caption{Uncertainties $\left. \Delta P_x\Delta P_y\right\vert _{\protect%
\psi _{n,m-n}}$ versus the generalized lower bound (a) for background fields 
$\protect\theta (t)=\protect\alpha e^{-\protect\gamma t}$ and $\Omega (t)=%
\protect\beta e^{\protect\gamma t}$ and (b) for background fields $\protect%
\theta (t)=\protect\alpha \sin (\protect\gamma t)$ and $\Omega (t)=\protect%
\beta \sin (\protect\gamma t/2)$. In both panels the constants are $\protect%
\alpha=5$, $\protect\beta=2$, $\protect\gamma=2$, $m=\hbar=\protect\tau=%
\protect\omega=1$, $\protect\kappa=1/4$ and $\protect\mu=\protect\sqrt{5/3}$.
}
\label{fig3}
\end{figure}

\begin{figure}[h]
\centering   \includegraphics[width=7.5cm,height=6.0cm]{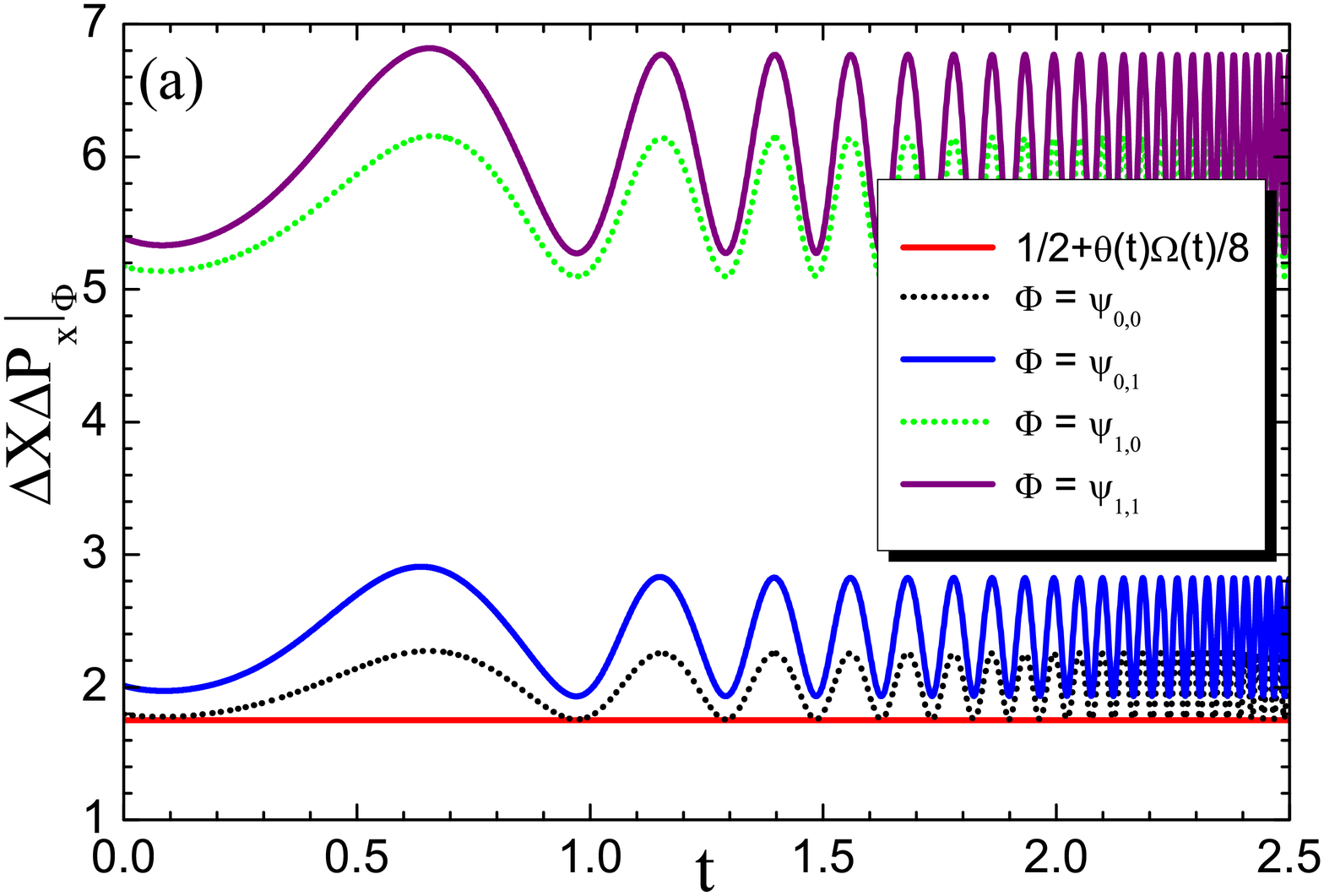} %
\includegraphics[width=7.5cm,height=6.0cm]{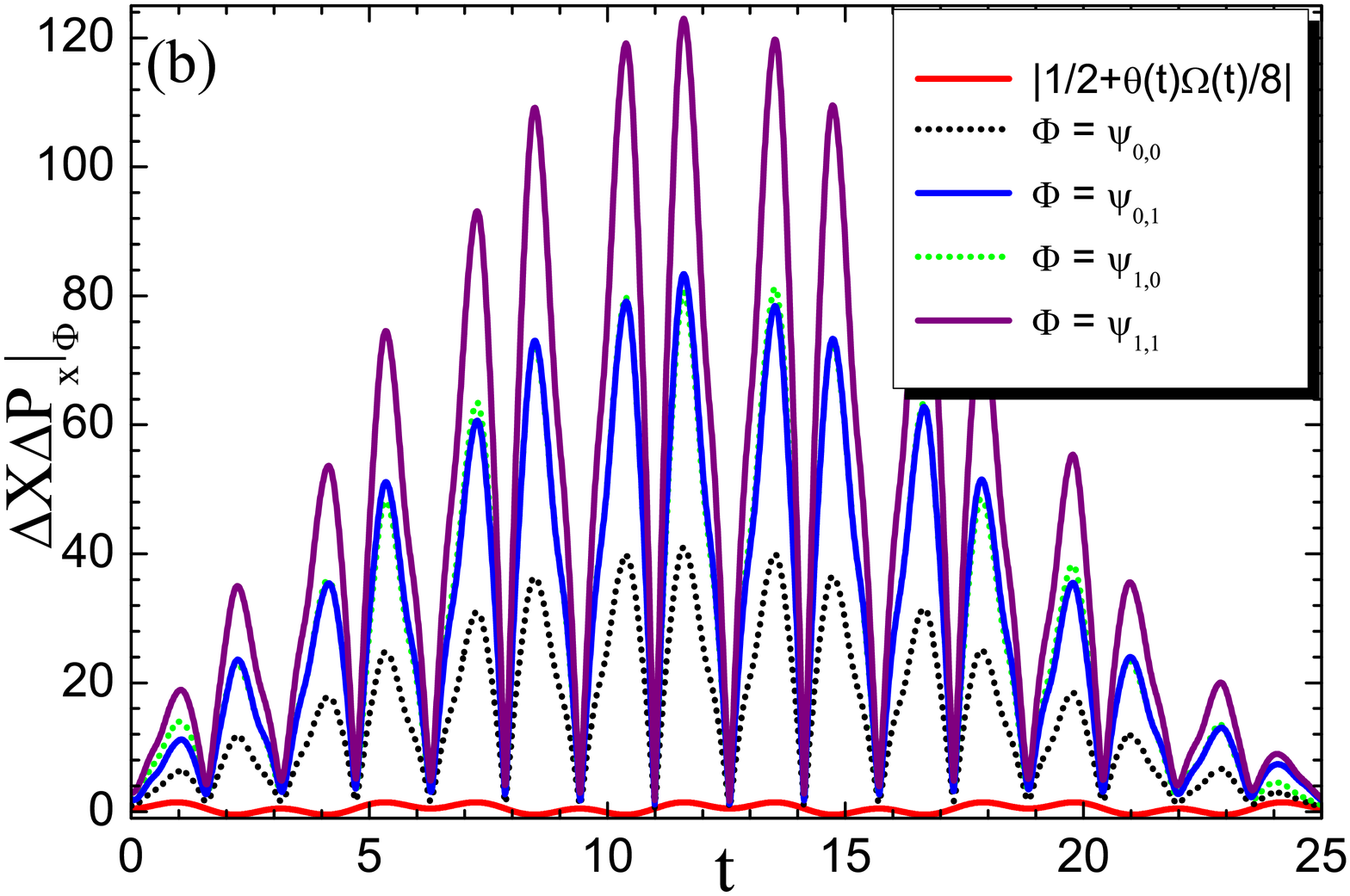} \centering   
\caption{Uncertainties $\left. \Delta X\Delta P_x\right\vert _{\protect\psi %
_{n,m-n}}$ versus the generalized lower bound (a) for background fields $%
\protect\theta (t)=\protect\alpha e^{-\protect\gamma t}$ and $\Omega (t)=%
\protect\beta e^{\protect\gamma t}$ and (b) for background fields $\protect%
\theta (t)=\protect\alpha \sin (\protect\gamma t)$ and $\Omega (t)=\protect%
\beta \sin (\protect\gamma t/2)$. In both panels the constants are $\protect%
\alpha=5$, $\protect\beta=2$, $\protect\gamma=2$, $m=\hbar=\protect\tau=%
\protect\omega=1$, $\protect\kappa=1/4$ and $\protect\mu=\protect\sqrt{5/3}$.
}
\label{fig4}
\end{figure}

\subsection{The generalized uncertainty relation for coherent states}

As is well known coherent states are convenient to use in a number of fields
of quantum theory, especially in quantum optics, because of the fact that by
definition they constitute the transition from a classical to a quantum
mechanical formulation of a given system. Starting with Schr{\"{o}}dinger's
investigations \cite{schrcs}, the first systematic and formal way was
developed by Glauber \cite{glauber}, who also coined the term coherent
states. Since some of properties are very specific to the harmonic
oscillator several types and generalizations of coherent states have been
proposed thereafter to accommodate different types of situations, see for
instance \cite{dodonov} for a review on the developments up to 2001. Fo
instance, so-called Klauder \cite{klauder0,klauder01} and Gazeau-Klauder 
\cite{gazeau_klauder} cherent states, for which the quantum classical
correspondence was recently investigated in \cite{klauder2,dey_fring_pra},
are extremely useful. 

Even though the model under consideration here is of course
not the harmonic oscillator, we still have the invariant $I(t)$ expressed in
terms of the time-dependent creation and annihilation operators. This
enables us to employ techniques used for the construction of Glauber
coherent states \cite{glauber}. Defining therefore the coherent states by
means of the time-dependent displacement operator $D(\alpha ,t)$ as 
\begin{equation}
\left\vert \alpha ,t\right\rangle :=D(\alpha ,t)\left\vert 0,0\right\rangle
,\quad \text{with \ }D(\alpha ,t):=e^{\alpha \hat{a}^{\dagger }(t)-\alpha
^{\ast }\hat{a}(t)},\quad 
\end{equation}%
it is immediately verified that they constitute eigenstates of the
annihilation operator $\hat{a}(t)$, i.e. $\hat{a}(t)\left\vert \alpha
,t\right\rangle =\alpha \left\vert \alpha ,t\right\rangle $. Using the
matrix elements for the expectation values with respect to the eigenfunction
(\ref{MA1})-(\ref{ma4}), we compute the expectation values with respect to
the Glauber coherent states 
\begin{eqnarray}
\left\langle \alpha ,t\right\vert x\left\vert \alpha ,t\right\rangle  &=&-%
\sqrt{\hbar }\sigma \func{Im}\alpha ,\quad \left\langle \alpha ,t\right\vert
x^{2}\left\vert \alpha ,t\right\rangle =\hbar \sigma ^{2}\left( \frac{1}{2}+%
\func{Im}^{2}\alpha \right) ,\quad  \\
\left\langle \alpha ,t\right\vert y\left\vert \alpha ,t\right\rangle  &=&-%
\sqrt{\hbar }\sigma \func{Re}\alpha ,\quad \left\langle \alpha ,t\right\vert
y^{2}\left\vert \alpha ,t\right\rangle =\hbar \sigma ^{2}\left( \frac{1}{2}+%
\func{Re}^{2}\alpha \right) ,\quad 
\end{eqnarray}

\begin{eqnarray}
\left\langle \alpha ,t\right\vert p_{x}\left\vert \alpha ,t\right\rangle  &=&%
\sqrt{\hbar }\left( \frac{\func{Re}\alpha }{\sigma }-\frac{\dot{\sigma}\func{%
Im}\alpha }{a}\right) ,\quad \left\langle \alpha ,t\right\vert
p_{x}^{2}\left\vert \alpha ,t\right\rangle =\frac{\hbar }{2}\left( \frac{1}{%
\sigma ^{2}}+\frac{\dot{\sigma}^{2}}{a^{2}}\right) +\left\langle \alpha
,t\right\vert p_{x}\left\vert \alpha ,t\right\rangle ^{2},  \notag \\
\left\langle \alpha ,t\right\vert p_{y}\left\vert \alpha ,t\right\rangle 
&=&-\sqrt{\hbar }\left( \frac{\func{Im}\alpha }{\sigma }+\frac{\dot{\sigma}%
\func{Re}\alpha }{a}\right) ,\quad \left\langle \alpha ,t\right\vert
p_{y}^{2}\left\vert \alpha ,t\right\rangle =\frac{\hbar }{2}\left( \frac{1}{%
\sigma ^{2}}+\frac{\dot{\sigma}^{2}}{a^{2}}\right) +\left\langle \alpha
,t\right\vert p_{y}\left\vert \alpha ,t\right\rangle ^{2},  \notag
\end{eqnarray}%
such that%
\begin{equation}
\left. \Delta x\right\vert _{\left\vert \alpha ,t\right\rangle }^{2}=\left.
\Delta y\right\vert _{\left\vert \alpha ,t\right\rangle }^{2}=\frac{\hbar
\sigma ^{2}}{2},\qquad \left. \Delta p_{x}\right\vert _{\left\vert \alpha
,t\right\rangle }^{2}=\left. \Delta p_{y}\right\vert _{\left\vert \alpha
,t\right\rangle }^{2}=\frac{\hbar }{2}\left( \frac{1}{\sigma ^{2}}+\frac{%
\dot{\sigma}^{2}}{a^{2}}\right) .  \label{UN1}
\end{equation}%
Notice that the uncertainties are the same as those computed with respect to
the ground state $\psi _{0,0}$. Likewise we compute%
\begin{equation}
\left. \Delta X\right\vert _{\left\vert \alpha ,t\right\rangle }^{2}=\left.
\Delta Y\right\vert _{\left\vert \alpha ,t\right\rangle }^{2}=\left. \Delta
X\right\vert _{\psi _{0,0}}^{2},\qquad \left. \Delta P_{x}\right\vert
_{\left\vert \alpha ,t\right\rangle }^{2}=\left. \Delta P_{y}\right\vert
_{\left\vert \alpha ,t\right\rangle }^{2}=\left. \Delta P_{x}\right\vert
_{\psi _{0,0}}^{2},\quad   \label{UN2}
\end{equation}%
such that the uncertainty relations are identical to those in (\ref{in1})-(%
\ref{in3}) with $\psi _{0,0}$ replaced by $\left\vert \alpha ,t\right\rangle 
$. The crucial difference is of course that $\psi _{0,0}$ is annihilated by $%
a(t)$, whereas $\left\vert \alpha ,t\right\rangle $ constitutes an
eigenstate for $\hat{a}(t)$.

Having creation and annihilation operators at our disposal we can use
standard techniques from quantum optics to construct squeezed states \cite%
{Knight} and improve on the uncertainties obtained so far. Employing for
this purpose the so-called squeezing operator $S(\beta ,t)$ by defining 
\begin{equation}
\left\vert \alpha ,\beta ,t\right\rangle :=S(\beta ,t)D(\alpha ,t)\left\vert
0,0\right\rangle ,\quad \text{with \ }S(\beta ,t):=e^{\frac{\beta }{2}[\hat{a%
}^{2}(t)-\hat{a}^{\dagger 2}(t)]},\quad 
\end{equation}%
we may compute the relevant matrix elements for these states, not reported
here. Using those we may subsequently deduce the uncertainties for the
auxiliary variables to%
\begin{eqnarray}
\left. \Delta x\right\vert _{\left\vert \alpha ,\beta ,t\right\rangle }^{2}
&=&\left. \Delta y\right\vert _{\left\vert \alpha ,-\beta ,t\right\rangle
}^{2}=\frac{\hbar }{2}\sigma ^{2}e^{\beta }\cosh \beta , \\
\left. \Delta p_{x}\right\vert _{\left\vert \alpha ,\beta ,t\right\rangle
}^{2} &=&\left. \Delta p_{y}\right\vert _{\left\vert \alpha ,-\beta
,t\right\rangle }^{2}=\frac{\hbar }{2}\left( \frac{1}{\sigma ^{2}}e^{-\beta
}+\frac{\dot{\sigma}^{2}}{a^{2}}e^{\beta }\right) \cosh \beta ,
\end{eqnarray}%
and for our noncommutative variables to%
\begin{eqnarray*}
\left. \Delta X\right\vert _{\left\vert \alpha ,\beta ,t\right\rangle }^{2}
&=&\left. \Delta Y\right\vert _{\left\vert \alpha ,-\beta ,t\right\rangle
}^{2}=\frac{\hbar }{2}\left[ \sigma ^{2}e^{\beta }+\frac{\theta ^{2}(t)}{%
4\hbar ^{2}}\left( \frac{1}{\sigma ^{2}}e^{\beta }+\frac{\dot{\sigma}^{2}}{%
a^{2}}e^{-\beta }\right) \right] \cosh \beta +\frac{\theta (t)}{4}%
(1-e^{2\beta }), \\
\left. \Delta P_{x}\right\vert _{\left\vert \alpha ,\beta ,t\right\rangle
}^{2} &=&\left. \Delta P_{y}\right\vert _{\left\vert \alpha ,-\beta
,t\right\rangle }^{2}=\frac{\hbar }{2}\left[ \frac{1}{\sigma ^{2}}e^{-\beta
}+\frac{\dot{\sigma}^{2}}{a^{2}}e^{\beta }+\frac{\Omega ^{2}(t)}{4\hbar ^{2}}%
\sigma ^{2}e^{-\beta }\right] \cosh \beta +\frac{\Omega (t)}{4}(1-e^{2\beta
}).
\end{eqnarray*}%
As expected these expressions reduce to (\ref{UN1}) and (\ref{UN2}) when $%
\beta \rightarrow 0$.

\begin{figure}[h]
\centering   \includegraphics[width=7.5cm,height=6.0cm]{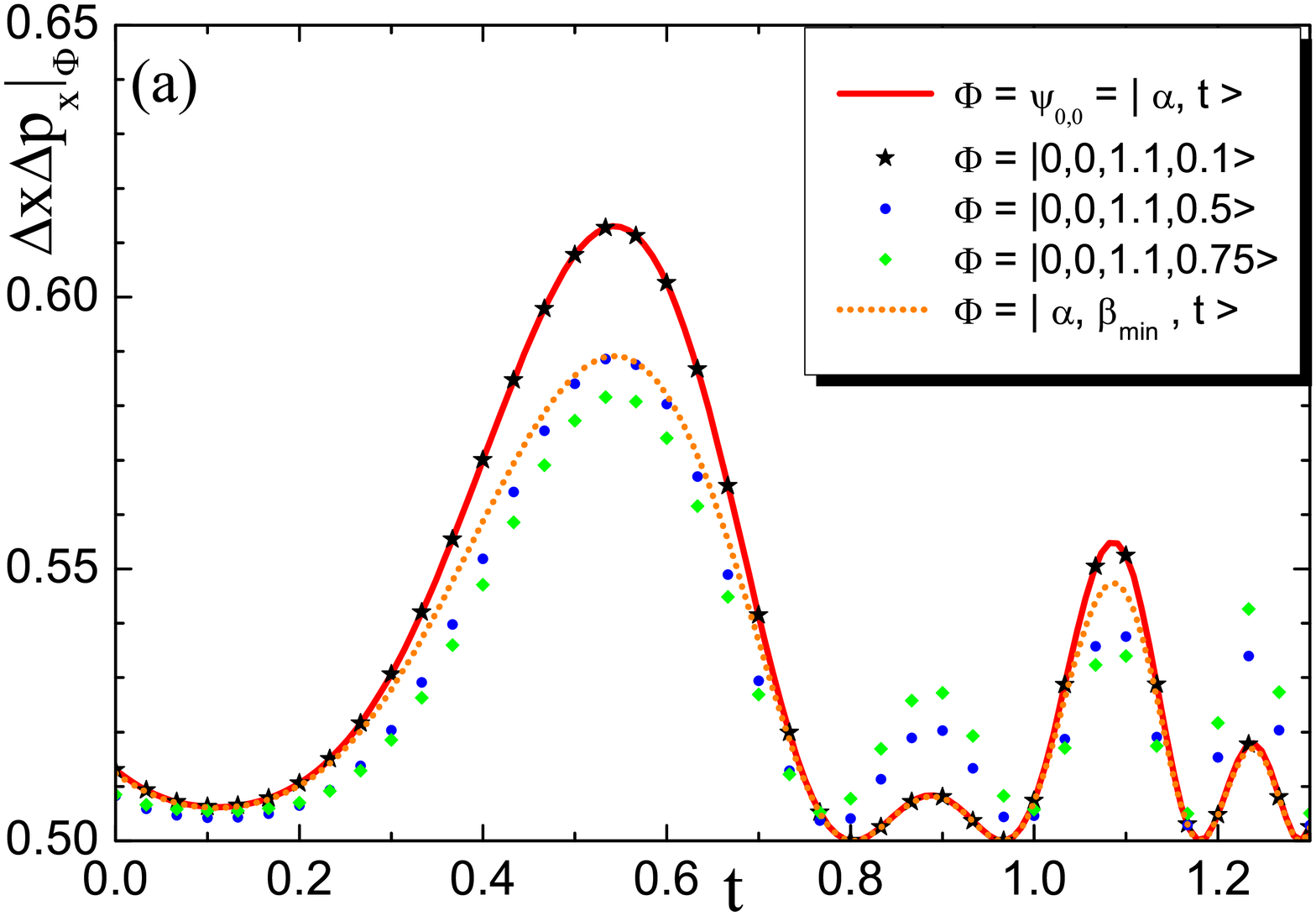} %
\includegraphics[width=7.5cm,height=6.0cm]{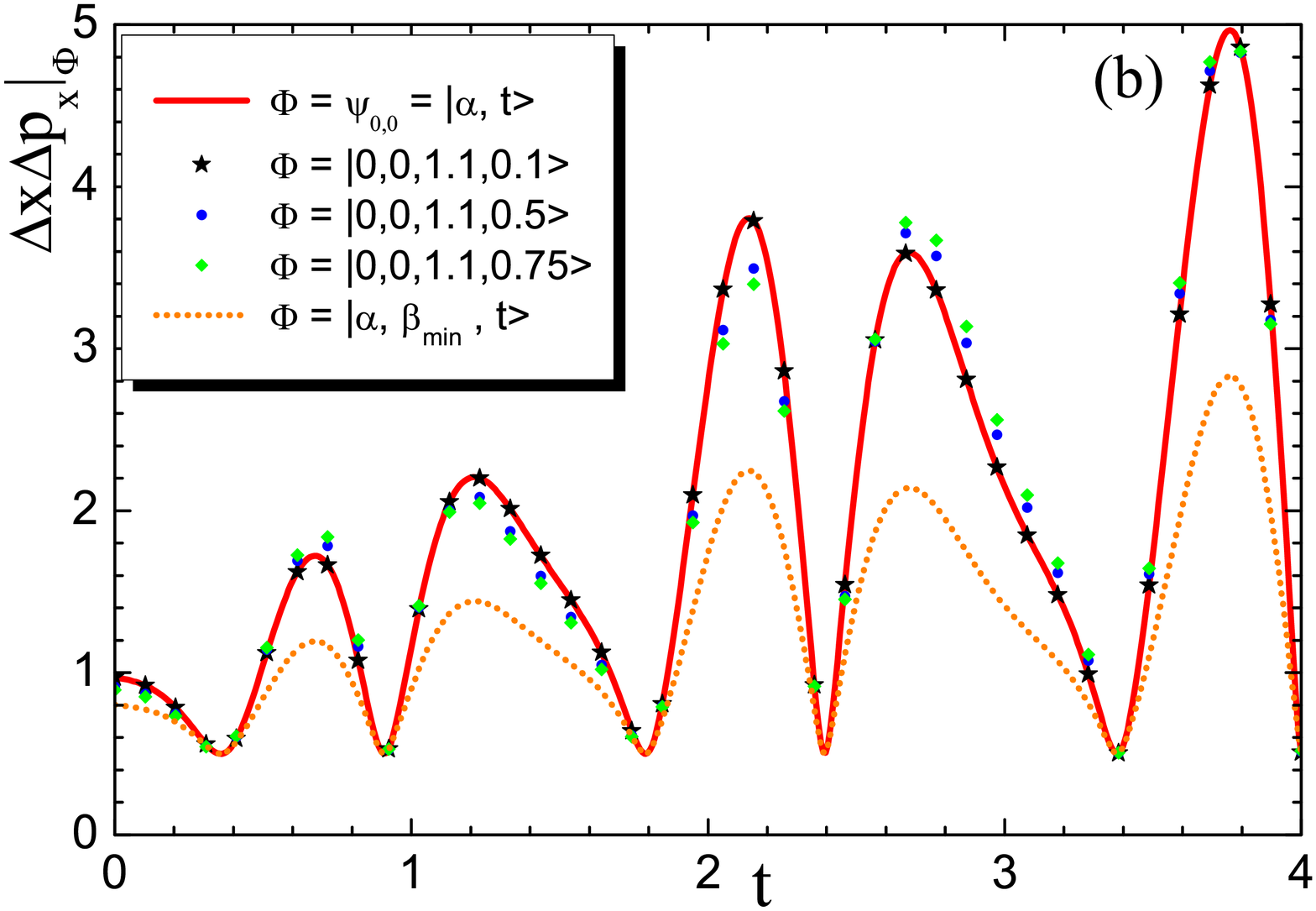} \centering   
\caption{Uncertainties with respect to Glauber coherent states versus
squeezed Glauber coherent states and Gaussian Klauder coherent states for
the auxiliary variables $x,p_{x}\,$, $\left. \Delta x\Delta p_{x}\right\vert
_{\left\vert \protect\alpha ,t\right\rangle }$ versus $\left. \Delta x\Delta
p_{x}\right\vert _{\left\vert \protect\alpha ,\protect\beta ,t\right\rangle
} $ versus $\left. \Delta x\Delta p_{x}\right\vert _{|GK>}$ (a) for
background fields $\protect\theta (t)=\protect\alpha e^{-\protect\gamma t}$
and $\Omega (t)=\protect\beta e^{\protect\gamma t}$ and (b) for background
fields $\protect\theta (t)=\protect\alpha \sin (\protect\gamma t)$ and $%
\Omega (t)=\protect\beta \sin (\protect\gamma t/2)$. In both panels the
constants are $\protect\alpha =5$, $\protect\beta =2$, $\protect\gamma =2$, $%
m=\hbar=\protect\tau=\protect\omega =1$, $\protect\kappa =1/4$ and $\protect%
\mu =\protect\sqrt{5/3}$.}
\label{fig5}
\end{figure}

We can now use the freedom to choose the function $\beta (t)$ to minimize
the uncertainties further. For instance, it is easily found that the
uncertainty $\Delta x\left. \Delta p_{x}\right\vert _{\left\vert \alpha
,\beta ,t\right\rangle }$ is minimal for $\beta (t)=\beta _{\min }(t)=1/2\ln %
\left[ \left( a\sqrt{a^{2}+8\sigma ^{2}\dot{\sigma}^{2}}-a^{2}\right)
/(4\sigma ^{2}\dot{\sigma}^{2})\right] $. Thus taking this value we should
find $\Delta x\left. \Delta p_{x}\right\vert _{\left\vert \alpha ,\beta
_{\min },t\right\rangle }<$ $\Delta x\left. \Delta p_{x}\right\vert
_{\left\vert \alpha ,t\right\rangle }$, which is indeed confirmed in figure %
\ref{fig5}, where we observe that squeezing leads to a considerable
reduction in the uncertainties.

\begin{figure}[h]
\centering   \includegraphics[width=7.5cm,height=6.0cm]{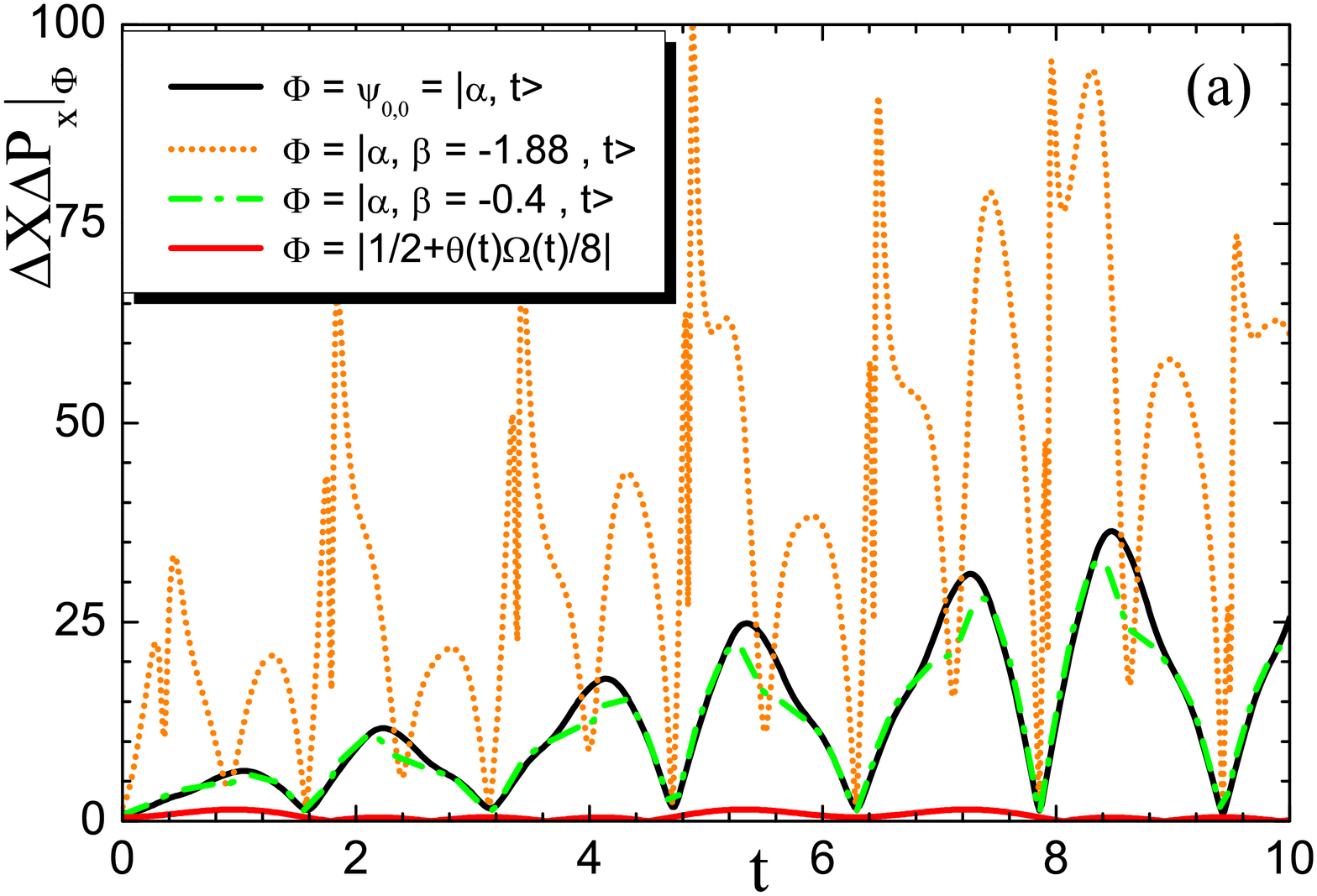} %
\includegraphics[width=7.5cm,height=6.0cm]{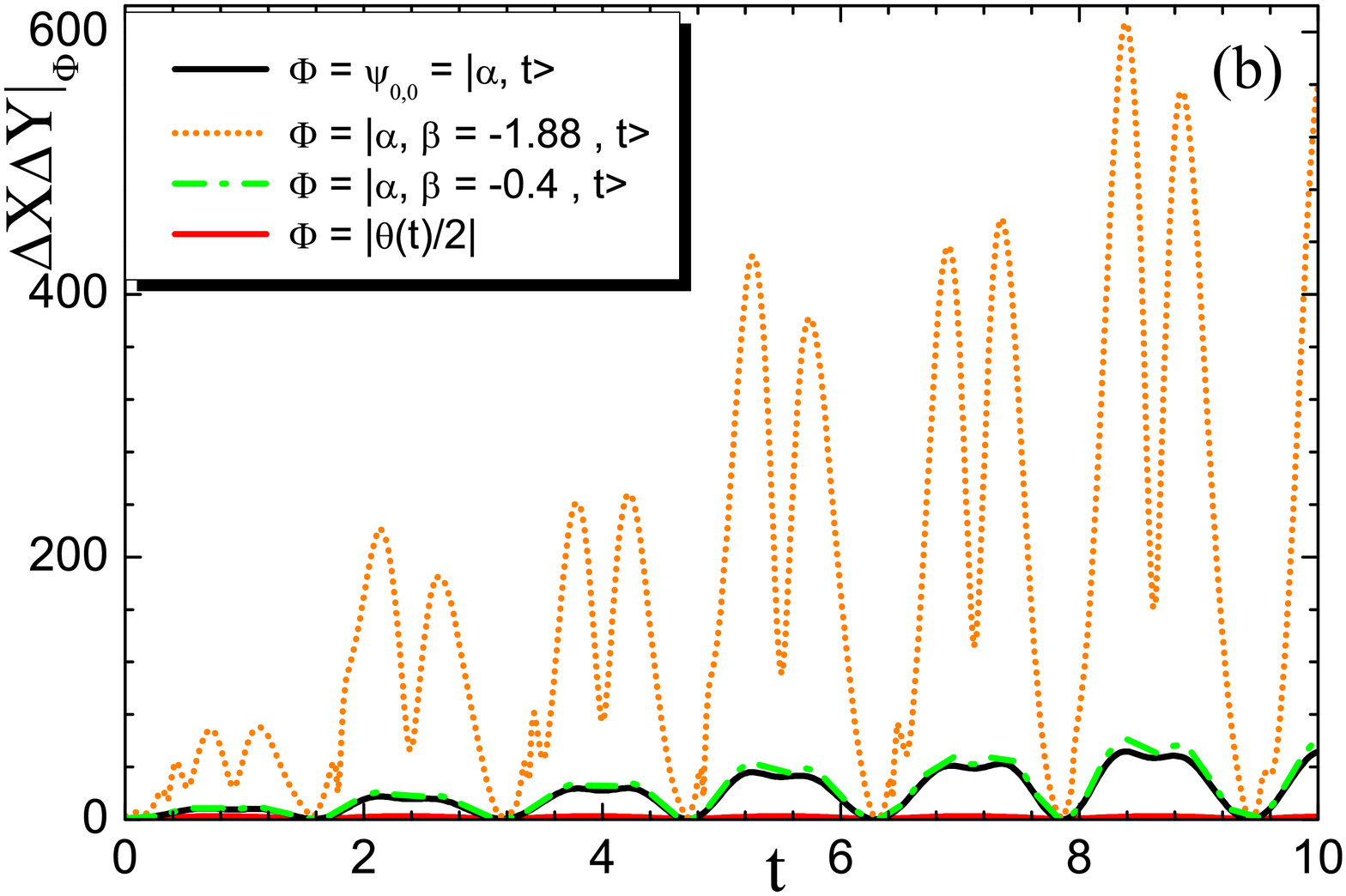} \centering   
\caption{ Uncertainties with respect to Glauber coherent states versus
squeezed Glauber coherent states for the noncommutative variables $%
X,Y,P_{x}\,$ for background fields $\protect\theta (t)=\protect\alpha \sin (%
\protect\gamma t)$ and $\Omega (t)=\protect\beta \sin (\protect\gamma t/2)$.
In both panels the constants are $\protect\alpha =5$, $\protect\beta =2$, $%
\protect\gamma =2$, $m=\hbar=\protect\tau=\protect\omega =1$, $\protect%
\kappa =1/4$ and $\protect\mu =\protect\sqrt{5/3}$.}
\label{fig6}
\end{figure}

The minimization for the uncertainties involving our noncommutative
variables is less obvious. Due to the complexity of the expressions we can
not perform this task for generic $\beta (t)$, but only for specific
instances in time. For instance, we find numerically the minimum for $\Delta
X\left. \Delta P_{x}\right\vert _{\left\vert \alpha ,\beta ,t=4\right\rangle
}$ at $\beta =-1.88203$. Indeed, as seen in figure \ref{fig6} panel (a), at $%
t=4$ this value leads to a reduction in the uncertainties when compared to $%
\Delta X\left. \Delta P_{x}\right\vert _{\left\vert \alpha ,t=4\right\rangle
}$.

 However, for different values of time the uncertainties have grown
considerably. It appears that the squeezing works only well for
momentum-coordinate uncertainties as for instance $\Delta X\left. \Delta
Y\right\vert _{\left\vert \alpha ,\beta ,t\right\rangle }$ is always minimal
at $\beta (t)=0$, such that the squeezing does not lead to any reduction in
these uncertainties. Figure \ref{fig6} panel (b) exhibits these findings.

Let us next compare our findings with the uncertainties computed with
respect to Gaussian Klauder coherent states defined as \cite%
{Fox,FoxChoi,Choig}%
\begin{equation}
\left\vert GK\right\rangle =\left\vert n,m_{0},\phi _{0},s\right\rangle :=%
\frac{1}{\sqrt{N(m_{0})}}\sum_{m=0}^{\infty }\exp \left[ -\frac{(m-m_{0})^{2}%
}{4s^{2}}\right] e^{im\phi _{0}}\left\vert n,m-n\right\rangle ,
\end{equation}%
with normalization factor $N(m_{0}):=\sum_{m=0}^{\infty }\exp \left[
-(m-m_{0})^{2}/(2s^{2})\right] $, initial phase factor $\phi _{0}$ and
Gaussian standard deviation $s$. Using the matrix elements (\ref{MA1})-(\ref%
{ma4}) we readily compute the expectation values with respect to these
states 
\begin{eqnarray}
\left\langle GK\right\vert x\left\vert GK\right\rangle  &=&-\frac{\sqrt{%
\hbar }}{N(m_{0})}\sigma \sin (\phi _{0}+\alpha _{01})S_{1}(m_{0}), \\
\left\langle GK\right\vert y\left\vert GK\right\rangle  &=&-\frac{\sqrt{%
\hbar }}{N(m_{0})}\sigma \cos (\phi _{0}+\alpha _{01})S_{1}(m_{0}), \\
\left\langle GK\right\vert p_{x}\left\vert GK\right\rangle  &=&\frac{\sqrt{%
\hbar }}{N(m_{0})}\left[ \frac{1}{\sigma }\cos (\phi _{0}+\alpha _{01})-%
\frac{\dot{\sigma}}{a}\sin (\phi _{0}+\alpha _{01})\right] S_{1}(m_{0}), \\
\left\langle GK\right\vert p_{y}\left\vert GK\right\rangle  &=&-\frac{\sqrt{%
\hbar }}{N(m_{0})}\left[ \frac{1}{\sigma }\sin (\phi _{0}+\alpha _{01})+%
\frac{\dot{\sigma}}{a}\cos (\phi _{0}+\alpha _{01})\right] S_{1}(m_{0}),
\end{eqnarray}%
and 
\begin{eqnarray}
\left\langle GK\right\vert x^{2},y^{2}\left\vert GK\right\rangle  &=&\frac{%
\hbar \sigma ^{2}}{2N(m_{0})}\left[ S_{2}(n+1,m_{0})\mp \sqrt{2}\cos (2\phi
_{0}+\alpha _{02})S_{3}(m_{0})\right] , \\
\left\langle GK\right\vert p_{x}^{2},p_{y}^{2}\left\vert GK\right\rangle  &=&%
\frac{\hbar }{2N(m_{0})}\left\{ \left( \frac{1}{\sigma ^{2}}+\frac{\dot{%
\sigma}^{2}}{a^{2}}\right) S_{2}(n+1,m_{0})\right.  \\
&&\left. \pm \sqrt{2}\left[ \left( \frac{1}{\sigma ^{2}}-\frac{\dot{\sigma}%
^{2}}{a^{2}}\right) \cos (2\phi _{0}+\alpha _{02})-2\frac{\dot{\sigma}}{%
a\sigma }\sin (2\phi _{0}+\alpha _{02})\right] S_{3}(m_{0})\right\} ,  \notag
\\
\left\langle GK\right\vert xp_{y},yp_{x}\left\vert GK\right\rangle  &=&\frac{%
\hbar }{2N(m_{0})}\left\{ \sqrt{2}\left[ \frac{\dot{\sigma}\sigma }{a}\sin
(2\phi _{0}+\alpha _{02})-\cos (2\phi _{0}+\alpha _{02})\right]
S_{3}(m_{0})\right.  \\
&&\left. \pm S_{2}(-n,m_{0})\right\} .  \notag
\end{eqnarray}%
We abbreviated $G(m,m_{0}):=\exp \left[ -(m-m_{0})^{2}/(4s^{2})\right] $ and
the sums%
\begin{eqnarray}
S_{1}(y) &:&=\sum_{k=0}^{\infty }\sqrt{k+1}G(k,y)G(k+1,y), \\
S_{2}(x,y) &:&=\sum_{k=0}^{\infty }(k+x)G^{2}(k,y), \\
S_{3}(y) &:&=\sum_{k=0}^{\infty }\mu (k,k+2)G(k,y)G(k+2,y).
\end{eqnarray}%
One could make some approximations here for the sums by replacing them with
Gaussian integrals, as for instance in \cite{FoxChoi,ChoiMM}. However, these
sums converge very fast with only some of the initial terms taken into
account and therefore it suffices here for our purposes to present numerical
values. When the Gaussian enveloping function is very sharp we notice that
the main contribution simply results from the center of the Gaussian. For
instance, for $s=0.1$, we compute $S_{1}(0)<10^{-10}$, $S_{2}(n,0)=n$, $%
S_{3}(0)<10^{-10}$ and $N(0)=1$, such that 
\begin{equation}
\left. \Delta o\right\vert _{\psi _{0,0}}^{2}=\left. \Delta o\right\vert
_{\left\vert \alpha ,t\right\rangle }^{2}=\left. \Delta o\right\vert
_{\left\vert GK\right\rangle }^{2}\quad \text{for }o=x,y,p_{x},p_{y}\text{.}
\end{equation}

This behaviour is clearly observable in figure \ref{fig5}. For a broader
Gaussian enveloping function other modes start to contribute. For instance,
for $s=0.5$ we compute $S_{1}(0)=0.3774$, $S_{2}(0,0)=0.1360$, $%
S_{2}(1,0)=1.2717$, $S_{3}(0)=0.0184$ and $N(0)=1.1357$ and for $s=0.75$ we
find $S_{1}(0)=0.7998$, $S_{2}(0,0)=1.9092$, $S_{2}(1,0)=0.4693$, $%
S_{3}(0)=0.1897$ and $N(0)=1.4400$. For these values the uncertainties for
the auxiliary variables are depicted in figure \ref{fig5} for two different
types of background fields. We observe that depending on the instance of
time the uncertainties might be lowered or increased.

When comparing with the uncertainties for the squeezed coherent states it
appears that optimal minimum is dependent on the type of background field.
We observe in figure \ref{fig5} that for sinusoidal background fields the
squeezed Glauber coherent states lead to minimal uncertainties which can not
be undercut when using Gaussian Klauder coherent states instead, whereas for
exponential backgrounds Gaussian Klauder coherent states allow for a further
minimization.

\section{Conclusions}

We have formulated and investigated a prototype model on a time-dependent
background. For an explicit representation of the underlying noncommutative
algebra the Hamiltonian naturally acquire a time-dependent form. Using the
Lewis-Riesenfeld method of invariants we constructed the time-dependent
invariants together with their eigensystem. Following the standard procedure
allowed to compute the eigenfunctions for the original Hamiltonian. As
common in the context of the invariant method all solutions are expressed in
terms the solutions of the nonlinear Ermakov-Pinney equation and variations
thereof. In general this auxiliary problem is not dealt with in this context
and all expressions are left as still dependent on an unknown function, $%
\sigma (t)$ in our case. In order to make the solutions more explicit and to
allow also for numerical studies thereafter, we have included here a
detailed discussion of some solutions.

Our explicit solutions then allow for a analysis of the generalized
uncertainty relations for which the lower bounds become time-dependent
functions. Since our invariants are expressed in terms of time-dependent
creation and annihilation operators, standard Glauber coherent states were
constructed by means of the displacement operator in a straightforward
manner. We found that the uncertainties for these states are identical to
those of the ground state annihilated by $a(t)$. By constructing the
so-called squeezing operator we demonstrated that these uncertainties can be
further minimized for momentum-coordinate uncertainties, where the absolute
lower bound was only be reached for certain instances in time. For
coordinate-coordinate uncertainties the minimal uncertainties were already
reached by the Glauber coherent states and squeezing does not lead to any
further improvement. We compared these findings with an analysis for
so-called Gaussian Klauder coherent states. A major difference towards the
forgoing computations is that the phase $\alpha _{n,\ell }(t)$ becomes a
relevant quantity. While in the computation of expectation values for
eigenstates the phase always cancels due to the sum in $\left\vert
GK\right\rangle $ it leads here to interferences. We observe that also for
the Gaussian Klauder coherent states the uncertainties resulting from the
computations for the ground state and the nonsqueezed Glauber coherent state
can be undercut. The answer to the question which type of the coherent
states is optimal appears to be background field dependent. The
time-dependent lowest bounds are well respected for all investigated
scenarios.

There remain a multitude of challenges. First of all it would be highly
desirable to investigate models on different types of time-dependent
backgrounds rather than (\ref{space}), possibly even those leading to
minimal length. As always the study of different types of models will
complete and enrich the understanding. The interesting question in all these
different types of scenarios is whether they still allow for explicit
solvability, which is one of the main virtue of our investigations, or if
one needs to resort to additional approximations.
\medskip 

\textbf{Acknowledgments:} SD is supported by a City University Research
Fellowship. AF thanks Abdelhafid Bounames and Boubakeur Khantoul for
discussions.


\end{document}